\documentclass[fleqn,usenatbib]{mnras}

\usepackage{newtxtext,newtxmath}
\usepackage{multirow}
\usepackage{tablefootnote}
\usepackage{threeparttable}

\usepackage[T1]{fontenc}

\DeclareRobustCommand{\VAN}[3]{#2}
\let\VANthebibliography\thebibliography
\def\thebibliography{\DeclareRobustCommand{\VAN}[3]{##3}\VANthebibliography}

\usepackage{graphicx}	
\usepackage{amsmath}	
\usepackage{subcaption}
\title[3D jet-CO disc orientation in nearby radio galaxies]{\centering{The AGN fuelling/feedback cycle in nearby radio galaxies \\
  III. 3D relative orientations of radio jets and CO discs and their interaction}}

\author[I. Ruffa et al.]{Ilaria Ruffa,$^{1,2}$\thanks{E-mail: i.ruffa@ira.inaf.it}
Robert A. Laing,$^{3}$
Isabella Prandoni,$^{2}$
Rosita Paladino,$^{2}$
Paola Parma,$^{2}$
\newauthor
Timothy A. Davis,$^{4}$
and Martin Bureau$^{5}$
\\
$^{1}$Dipartimento di Fisica e Astronomia, Universit\`{a} degli Studi di Bologna, via P.\ Gobetti 93/2, 40129, Bologna, IT\\
$^{2}$INAF - Istituto di Radioastronomia, via P.\ Gobetti 101, 40129, Bologna, IT\\
$^{3}$Square Kilometre Array Organisation, Jodrell Bank Observatory, Lower Withington, Macclesfield, Cheshire SK11 9FT, UK\\
$^{4}$School of Physics \& Astronomy, Cardiff University, Queens Buildings, The Parade, Cardiff CF24 3AA, UK\\
$^{5}$Sub-dept.\ of Astrophysics, Dept.\ of Physics, University of Oxford, Denys Wilkinson Building, Keble Road, Oxford OX1 3RH, UK
}

\date{Accepted XXX. Received YYY; in original form ZZZ}

\pubyear{2020}

\begin{document}
\label{firstpage}
\pagerange{\pageref{firstpage}--\pageref{lastpage}}
\maketitle

\begin{abstract}
This is the third paper of a series exploring the multi-frequency properties of a sample of eleven nearby low-excitation radio galaxies (LERGs) in the southern sky. We are conducting an extensive study of different galaxy components (stars, dust, warm and cold gas, radio jets) with the aim of better understanding the AGN fuelling/feedback cycle in LERGs.
Here we present new, deep, sub-kpc resolution Karl G.\,Jansky Very Large Array (JVLA) data for five sample sources at 10~GHz. Coupling these data with previously-acquired Atacama Large Millimeter/submillimeter Array (ALMA) CO(2-1) observations and measurements of comparable quality from the literature, we carry out for the first time a full 3D analysis of the relative orientations of jet and disc rotation axes in six FR\,I LERGs. This analysis shows (albeit with significant uncertainties) that the relative orientation angles span a wide range ($\approx$30$^{\circ}-60^{\circ}$). There is no case where both axes are accurately aligned and there is a marginally significant tendency for jets to avoid the disc plane. Our study also provides further evidence for the presence of a jet-CO disc interaction (already inferred from other observational indicators) in at least one source, NGC\,3100. In this case, the limited extent of the radio jets, along with distortions in both the molecular gas and the jet components, suggest that the jets are young, interacting with the surrounding matter and rapidly decelerating.
\end{abstract}




\section{Introduction}\label{sec:intro}
Evidence has accumulated that the large amount of energy deposited by feedback processes from active galactic nuclei (AGN) in the surrounding environment can have a strong impact on the subsequent evolution of the host galaxy, by - for example - significantly altering the physics and distribution of the inter-stellar medium (ISM) and thus modifying the star formation processes \citep[e.g.][]{Garcia14,Combes17,Harrison17,Harrison18}. AGN feedback is commonly invoked in two (non-exclusive) forms, each associated with a fundamental type of nuclear activity: radiative (or quasar) mode and kinetic (or jet) mode \citep[e.g.][]{Heckman14,Morganti17,Combes17}. The former occurs when the dominant energetic output from the nuclear activity is in the form of electromagnetic radiation (\citealp[e.g.][]{Wagner16,Bieri17}) and is typically associated with (quasar- or Seyfert-like) AGN accreting matter at high rates ($\ga 0.01$ \.{M}$_{\rm Edd}$, where \.{M}$_{\rm Edd}$ is the Eddington accretion rate\footnote{$\dot{M}_{\rm Edd} = \dfrac{4\pi \, G \, M_{\rm SMBH} \, m_{\rm p}}{\varepsilon \, c \, \sigma_{\rm T}}$, where $G$ is the gravitational constant, $M_{\rm SMBH}$ is the mass of the central super-massive black hole, $m_{\rm p}$ is the mass of the proton, $\varepsilon$ is the accretion efficiency, $c$ is the speed of light and $\sigma_{\rm T}$ is the cross-section for Thomson scattering.}). Kinetic (jet-mode) feedback instead occurs when the bulk of the energy generated from the accretion process is channelled into collimated outflows of non-thermal plasma (i.e.\,the radio jets) and is typically (but not exclusively) associated with low accretion rates \citep[$\ll 0.01$ \.{M}$_{\rm Edd}$; e.g.][]{Cielo18}. 

On large (i.e.\,hundreds of kpc) scales, radio jets produce one of the most spectacular manifestations of AGN feedback, that is the inflation of cavities in the hot X-ray emitting gas. This phenomenon provides the most striking evidence that expanding radio jets from massive radio galaxies at the centre of groups and clusters heat the surroundings, balancing radiative losses and thus preventing the gas from further cooling \citep[e.g.][]{McNamara12}. 

Kinetic AGN feedback, however, has also been observed to affect the host galaxy environment on much smaller (kpc or sub-kpc) scales. Indeed, jet-driven multi-phase gas outflows have been imaged in a number of galaxies (mostly Seyferts) hosting (sub-)kpc scale jets (\citealp[e.g.][]{Alatalo11,Combes13,Garcia14,Morganti15,Mahony16,Zovaro19,Murthy19}), showing that the distribution, kinematics and physical conditions of the galactic gaseous reservoirs can be radically altered by their interaction with the jets. These observational findings are consistent with 3D hydrodynamical simulations \citep[e.g.][]{Wagner12,Wagner16,Mukherjee16,Mukherjee18a,Mukherjee18b}, which demonstrate that jets expanding through the surrounding medium produce turbulent cocoons of shocked gas that can be accelerated up to high velocities ($>1000$~km~s$^{-1}$) over a wide range of directions. Such simulations also show the interaction processes to be very sensitive to both the jet power and the relative orientation of jets and gas discs \citep[e.g.][]{Mukherjee16,Mukherjee18a,Mukherjee18b}. In this scenario, jets of intermediate power (P$_{\rm jet}\sim10^{43}-10^{44}$
~erg~s$^{-1}$) at large ($\geq45^{\circ}$) angles to the disc are predicted to interact more strongly and for longer times than jets of higher power (P$_{\rm jet}>10^{44}$
~erg~s$^{-1}$) oriented perpendicularly to the disc plane. These theoretical studies, however, mostly focus on jets with powers in the range $10^{44} - 10^{46}$~erg~s$^{-1}$, typical of luminous radio galaxies with Fanaroff-Riley type II morphologies \citep[FR\,II; ][]{Fanaroff74}. The impact of jets of lower powers (more often associated with low-luminosity FR\,I radio galaxies; \citealp[e.g.][]{Godfrey13}) on the host-galaxy ISM is not yet clear.

In our project, we are interested in gaining a better understanding of the galaxy-scale fuelling/feedback cycle in a specific class of low-power radio galaxies, known as low-excitation radio galaxies (LERGs).  LERGs are known to be the numerically dominant radio galaxy population in the local Universe \citep[e.g.][]{Hardcastle07} and are predominantly hosted by red, massive ($M_{*} \geq 10^{11}~M_{\odot}$) early-type galaxies (ETGs; \citealp[e.g.][]{Best12}). These objects mostly have FR\,I radio morphologies and low or moderate radio luminosities at 1.4\,GHz ($P_{\rm 1.4GHz}<10^{25}$~W~Hz$^{-1}$). The central super-massive black holes (SMBHs) in LERGs 
accrete matter at low rates ($\ll 0.01$ \.{M}$_{\rm Edd}$) and produce almost exclusively kinetic feedback \citep[see e.g.][for a review]{Heckman14}. Despite their prevalence, systematic spatially-resolved studies of LERGs are very few so far, thus their trigger mechanisms and associated AGN feedback processes are still poorly understood \citep[e.g.][]{Hardcastle18}: investigating the nature of LERGs is crucial to shed light on the mechanisms which determine the observed properties of massive nearby ETGs. 

We are carrying out an extensive study of various galaxy components (stars, warm and cold gas, dust, radio jets) in a complete volume and flux-limited ($z<0.03$, $S_{2.7 \rm GHz}\geq 0.25$~Jy) sample of eleven LERGs in the southern sky. In \citet[][hereafter Paper I]{Ruffa19a} we presented ALMA Cycle 3 CO(2-1) and 230~GHz continuum observations of nine sample members. Our work shows that rotating (sub-)kpc molecular discs are very common in LERGs. Accurate 3D modelling of these discs \citep[][hereafter Paper II]{Ruffa19b} demonstrates that, although the bulk of the gas is in ordered rotation (at least at the resolution of our ALMA observations), low-amplitude perturbations and/or non-circular motions are ubiquitous. Such asymmetries indicate that the gas is not fully relaxed into the host galaxy potential and are strongly suggestive of the presence of radio jet--CO disc interactions in at least two cases. An analysis of the relative orientations of jets and discs derived from spatially resolved imaging can provide further clues in this regard. 

Many authors have studied the relative orientation of jets and the axes of dust discs or lanes {\em as they appear projected onto the plane of the sky}  (\citealt*{KE79,Moell92,vDF95,deKoff00}; \citealt{deRuiter02}), finding a significant preference for alignment, but with clear exceptions.  \citet{Schmitt02} and \citet{Verdoes05} pointed out that the aligned cases often exhibit irregular, patchy dust {\em lanes}, but that well-defined dust {\em discs} show a wider range of orientations with respect to the jets. The gas discs we have found in our sample are associated with regular dust discs. Comparing our ALMA CO observations with archival radio data (Paper~I), we find a range of {\em projected} misalignment angles consistent with the aforementioned earlier works.  Projection, however,  significantly affects the observed distribution of relative orientation angles, so that a full 3D analysis is crucial to draw physically meaningful conclusions. \citet{Schmitt02} and \citet{Verdoes05} first attempted to carry out a statistical analysis of the problem in 3D by estimating the distributions of intrinsic dust disc -- jet misalignment angles ($\theta_{\rm dj}$) for small samples of nearby (mostly FR\,I) RGs, starting from the projected disc-jet position angle differences, $\Delta$, and the inclinations of the dust discs inferred from their ellipticities.  They found broad ranges of $\theta_{\rm dj}$, consistent with a homogeneous distribution over $0 \la \theta_{\rm dj} \la 55^\circ - 77^\circ$ \citep{Schmitt02} and an isotropic distribution \citep{Verdoes05}, respectively. Analyses of this type, however, can only constrain the distribution of misalignment angles for an ensemble of galaxies. Our aim here is to estimate the misalignment angles for individual objects by adding jet inclinations from radio imaging. This allows us to derive the relative orientations of jet and disc rotation axes in three dimensions. In addition, fitting the velocity fields of the CO discs gives us precise values for their inclinations, avoiding the need to assume that dust discs are circular.  

In this paper, we present high-resolution Karl G.\,Jansky Very Large Array (JVLA) continuum observations at 10~GHz for the sub-sample of our sources that are suitable for a study of the relative orientation between the jets and CO discs. The combination of these new deep radio data with the ALMA CO observations from Papers I and II, and the addition of data of comparable quality from the literature allows us to perform {\em for the first time} a full 3D analysis for six low-power FR\,I LERGs.

The paper is structured as follows. In Section~\ref{sec:obs} we describe the JVLA observations and data reduction. We present the modelling of the radio jets  in Section~\ref{sec:method}. We discuss the results in Section~\ref{sec:results}, before summarising and concluding in Section~\ref{sec:conclusion}. A detailed mathematical description of the 3D jet-disc system is provided in Appendix~\ref{sec:Appendix}.

Throughout this work we assume a standard $\Lambda$CDM cosmology with H$_{\rm 0}=70$\,km\,s$^{-1}$\,Mpc$^{\rm -1}$, $\Omega_{\rm \Lambda}=0.7$ and $\Omega_{\rm M}=0.3$. 

\section{Southern radio galaxy sample}\label{sec:obs}
\subsection{Sample description and ALMA observations}
Full details on the sample selection and ALMA observations can be found in Paper I; a brief summary is presented here.

Starting from the radio flux- and optical apparent magnitude-limited ($S_{2.7 \rm GHz}\geq 0.25$~Jy, m\textsubscript{v}$<17.0$) southern ($-17^{\circ}<\delta<-40^{\circ}$) radio galaxy sample of \citet{Ekers89}, we selected those sources satisfying the following criteria:
\begin{itemize}
\item elliptical/S0 galaxy optical counterpart;
\item host galaxy redshift $z<0.03$.
\end{itemize}
This resulted in a complete volume-limited sub-sample of eleven RGs, all with low or intermediate 1.4 GHz radio luminosities (P$_{\rm 1.4GHz}\leq10^{25.5}$~W~Hz$^{-1}$) and most with FR\,I radio morphologies. Based on the available optical spectroscopy (\citealt{Tad93,Smith00,Colless03,Coll06,Jones09}), all of the radio galaxies in this sample have [OIII] line luminosities below the relation shown in Figure 2 of \citet{Best12} and, as argued in that paper, can be securely classified as LERGs. 

Nine sample members were observed during ALMA Cycle 3 in the $^{12}$CO(2-1) line and 230~GHz continuum at spatial resolutions of a few hundred parsecs. Six out of nine sources observed with ALMA were detected in CO, with typical molecular gas masses ranging from $\approx10^{7}$ to $\approx10^{8}$~M$\odot$. 

We acquired new JVLA observations for a subset of the sample sources observed with ALMA with the primary aim of estimating the jet inclinations. This requires sensitive radio continuum data at high spatial resolution (to image the jet structure close to the nucleus; see Section~\ref{sec:method}).
Specifically, from the nine sources observed in $^{12}$CO(2-1), we selected those that meet the following criteria:
\begin{itemize}
\item resolved $^{12}$CO(2-1) emission detected with ALMA; 
\item images from archival VLA data showing sub-kpc scale jet structure.
\end{itemize}

Five of the six sources detected in $^{12}$CO(2-1) were found to satisfy 
these criteria:  IC\,1531, NGC\,3100, NGC\,3557, IC\,4296 and NGC\,7075 (NGC\,612 was excluded as we found no evidence for small-scale jet structure). Four of these sources show straight FR\,I jet structures, allowing us to estimate jet inclinations and, in turn, intrinsic jet--CO disc orientations as described in Section~\ref{sec:method}. The exception, NGC\,3100, has a distorted radio structure, but is of interest for a different reason: it shows convincing evidence for a jet--CO disc interaction (see Papers I and II) and the new, high-resolution JVLA observations would allow us to probe this in greater detail. To improve the statistics of jet--disc orientations, we have also extended our analysis to two other FR\,I LERGs for which CO and jet inclination data of comparable quality are available in the literature, NGC\,383 and NGC\,3665. 

A summary of the general (radio and CO) properties of all the sources analysed in this paper is presented in Table~\ref{tab:Southern Sample}.

\begin{table*}
\centering
\caption{General properties and main parameters of the CO discs of the radio galaxies analysed in this paper.}\label{tab:Southern Sample}
\begin{tabular}{l l c c c r c r r}
\hline
\multicolumn{1}{c}{ Radio} &
\multicolumn{1}{c}{ Host } &
\multicolumn{1}{c}{ z } & 
\multicolumn{1}{c}{ log~P$_{\rm 1.4GHz}$ } &
\multicolumn{1}{c}{ D$_{\rm L}$ } &
\multicolumn{3}{c}{ CO disc parameters} &
\multicolumn{1}{c}{  } \\
\multicolumn{1}{c}{ source } & 
\multicolumn{1}{c}{ galaxy } &
\multicolumn{1}{c}{ }  &  
\multicolumn{1}{c}{ } &
\multicolumn{1}{c}{ }  &
\multicolumn{1}{c}{maj. axis  $\times$ min. axis } &
\multicolumn{1}{c}{PA$_{\rm kin}$ }   &
\multicolumn{1}{c}{ M\textsubscript{mol}}   \\
\multicolumn{1}{c}{ } &
\multicolumn{1}{c}{ } &
\multicolumn{1}{c}{  } &
\multicolumn{1}{c}{ (W\,Hz$^{-1}$) } &
\multicolumn{1}{c}{ (Mpc)}  &
\multicolumn{1}{c}{ (pc $\times$ pc)} &
\multicolumn{1}{c}{ (deg)} &
\multicolumn{1}{c}{ (M$_{\rm \odot}$) } \\
\multicolumn{1}{c}{(1)} &
\multicolumn{1}{c}{ (2)} &
\multicolumn{1}{c}{ (3)} &
\multicolumn{1}{c}{ (4)} &
\multicolumn{1}{c}{ (5) } &
\multicolumn{1}{c}{ (6) } &
\multicolumn{1}{c}{ (7)} &
\multicolumn{1}{c}{ (8)} \\
\hline
PKS 0007$-$325&  IC\,1531 & 0.0256 & 23.9 & 112.0  & (250$\pm$50)\,$\times$\,(220$\pm$60) & 356.0$\pm1.0$ & $(1.1\pm0.1)\times10^{8}$\\
PKS 0958$-$314& NGC\,3100 & 0.0088 & 23.0 & 38.0  & (1600$\pm$300)\,$\times$\,(500$\pm$80) & 220.0$\pm0.1$ & $(1.2\pm0.1)\times10^{8}$\\
PKS 1107$-$372& NGC\,3557 & 0.0103 & 23.3 & 44.5 & (300$\pm$20)\,$\times$\,(200$\pm$10) & $211.0\pm1.0$ & $(6.2\pm0.6)\times10^{7}$ \\
PKS 1333$-$33 &  IC\,4296 & 0.0125 & 24.2 & 53.9 &  (200$\pm$20)\,$\times$\;\;(40$\pm$20) & 230.0$\pm$3.0 & $(2.0\pm0.2)\times10^{7}$ \\
PKS 2128$-$388& NGC\,7075 & 0.0185 & 23.8 & 80.3 & $<200\;\;\;\;\;\;\;\;\;\;\;\;$ & 322.0$\pm4.0$ & $(2.9\pm0.2)\times10^{7}$  \\
\hline
\multicolumn{9}{c}{\small{\textbf{Literature data}}}\\
\hline
3C31 & NGC\,383 & 0.0169 & 24.5 & 73.3 & 1400\,$\times$\,1600\;\;\;\;\;\; & 142.2 & $(2.1\pm0.2)\times10^{9}$\\
B2\,1122+39 &  NGC\,3665 & 0.0067 & 22.0 & 28.8 & 1600\,$\times$\,400\;\;\;\;\;\;\; & 206.0 & $(8.1\pm0.1)\times10^{8}$  \\ 
\hline 
\end{tabular}
\parbox[t]{1\textwidth}{ \textit{Notes.} Columns: (1) Name of the radio source. (2) Host galaxy name. (3) Galaxy redshift taken from the NASA/IPAC extragalactic database (NED). (4) Radio power at 1.4~GHz derived from the most accurate 1.4~GHz radio flux density given in NED (including all the radio emission associated with the source). (5) Luminosity distance derived from the redshift given in column (3). (6) CO disc dimensions. For the southern radio galaxies these are the disc major and minor axes (FWHM) deconvolved from the synthesized beam, as derived in Paper~I; for NGC\,383 and NGC\,3665 the axis lengths are approximate deconvolved sizes derived from the CO data presented in \citealp{North19} and \citealp{Onishi17}, respectively. (7) Kinematic position angle of the CO disc, i.e.\,orientation angle projected onto the plane
of the sky of the CO disc major axis, measured anticlockwise from North to the redshifted side of the CO velocity field and ranged $0 - 360^{\circ}$. These are best-fit position angles from the 3D kinematical modelling of the CO discs presented in Paper II (southern sample), \citealp{North19} (NGC\,383) and \citealp{Onishi17} (NGC\,3665).
(8) Molecular gas mass as derived in Paper I. For NGC\,383 and NGC\,3665 the molecular gas masses are estimated from the CO data presented in \citealp{North19} and \citealp{Alatalo13}, respectively.
All calculations assume the cosmology given in Section~\ref{sec:intro}.}
\end{table*}

\begin{table*}
\centering
\caption{Details of the X-band JVLA observations.}
\label{tab:JVLA obs}
\begin{tabular}{l r c c c r r}
\hline
\multicolumn{1}{c}{ Target } &
\multicolumn{1}{c}{ Date } & 
\multicolumn{1}{c}{ Time } & 
\multicolumn{1}{c}{   $\theta$\textsubscript{maj} } & 
\multicolumn{1}{c}{   $\theta$\textsubscript{min}  } & 
\multicolumn{1}{c}{   PA$_{\rm beam}$} & 
\multicolumn{1}{c}{   Scale }\\        
\multicolumn{1}{c}{  } &       
\multicolumn{1}{c}{  } &   
\multicolumn{1}{c}{   (min)} &          
\multicolumn{2}{c}{ (arcsec) } &   
\multicolumn{1}{c}{   (deg) } &
\multicolumn{1}{c}{   (pc)} \\       
\multicolumn{1}{c}{   (1) } &   
\multicolumn{1}{c}{   (2) } &
\multicolumn{1}{c}{   (3) } &
\multicolumn{1}{c}{   (4) } & 
\multicolumn{1}{c}{   (5) } &   
\multicolumn{1}{c}{   (6) } &
\multicolumn{1}{c}{   (7) } \\
\hline
 IC\,1531 &  2018-03-03 & 72 &    0.43  &  0.13  & $-19.6$  & 220  \\
 NGC\,3100 &    2018-03-24    &   66  &        0.37    &    0.14   &    6.41   &    70   \\
  NGC\,3557  &    2018-03-24    &    66   &     0.47   &    0.12    &    0.70    &    100    \\ 
 IC\,4296  &    2018-03-24   &   66  &        0.43    &     0.14    &    $-15.2$    &    110 \\  
 NGC\,7075 &    2018-03-03     &    72 &    0.46   &     0.13   &    0.62     &    180   \\
\hline
\end{tabular}
\parbox[t]{1\textwidth}{ \textit{Notes.} $-$ Columns: (1) Target name. (2) Observation date. (3) Total integration time on-source. (4) Major-axis FWHM of the synthesized beam. (5) Minor-axis FWHM of the synthesized beam. (6) Position angle of the synthesized beam. (7) Spatial scale corresponding to the major axis FWHM of the synthesized beam.}
\end{table*}

\subsection{JVLA observations}
We used the JVLA in A configuration to observe the five sample sources at X-band (centre frequency 10~GHz) in two observing blocks. Table~\ref{tab:JVLA obs} summarises the details of the observations (project code: 18A-200; PI: I.\,Ruffa). 
In both observing blocks the same spectral configuration was used: the frequency range of the X-band receiver (8-12~GHz) was divided into 32 spectral windows with 64 channels each and a channel width of 2~MHz. 3C48 and 3C286 were used as primary flux calibrators; J2109-4110, J2359-3133, J1037-2934, J1147-3812 and J1316-3338 were used as phase calibrators. Polarization data were also acquired and will be presented in a future paper.

The data were reduced using the Common Astronomy Software Applications (\textsc{CASA}) pipeline (version 5.4.1), which automatically processes each observing block by performing basic flagging and calibration. After the pipeline processing, we carefully inspected the calibrated data for the presence of residual radio-frequency interference (RFI) signal. The first observing block showed higher RFI levels than the second and required additional flagging. We used the \textsc{CASA} task \texttt{flagdata} in the auto-flagging mode \texttt{rflag} to first identify and then crop out the outlier signal in the time-frequency plane. In total, about 45\% of the data were flagged using this strategy. No significant additional flagging was needed in the second observing block.

\subsubsection{Imaging}
\begin{figure*}
\centering
\begin{subfigure}[t]{.45\textwidth}
\centering
\caption{\textbf{IC\,1531}}\label{fig:ic1531_VLA}
\includegraphics[width=\linewidth]{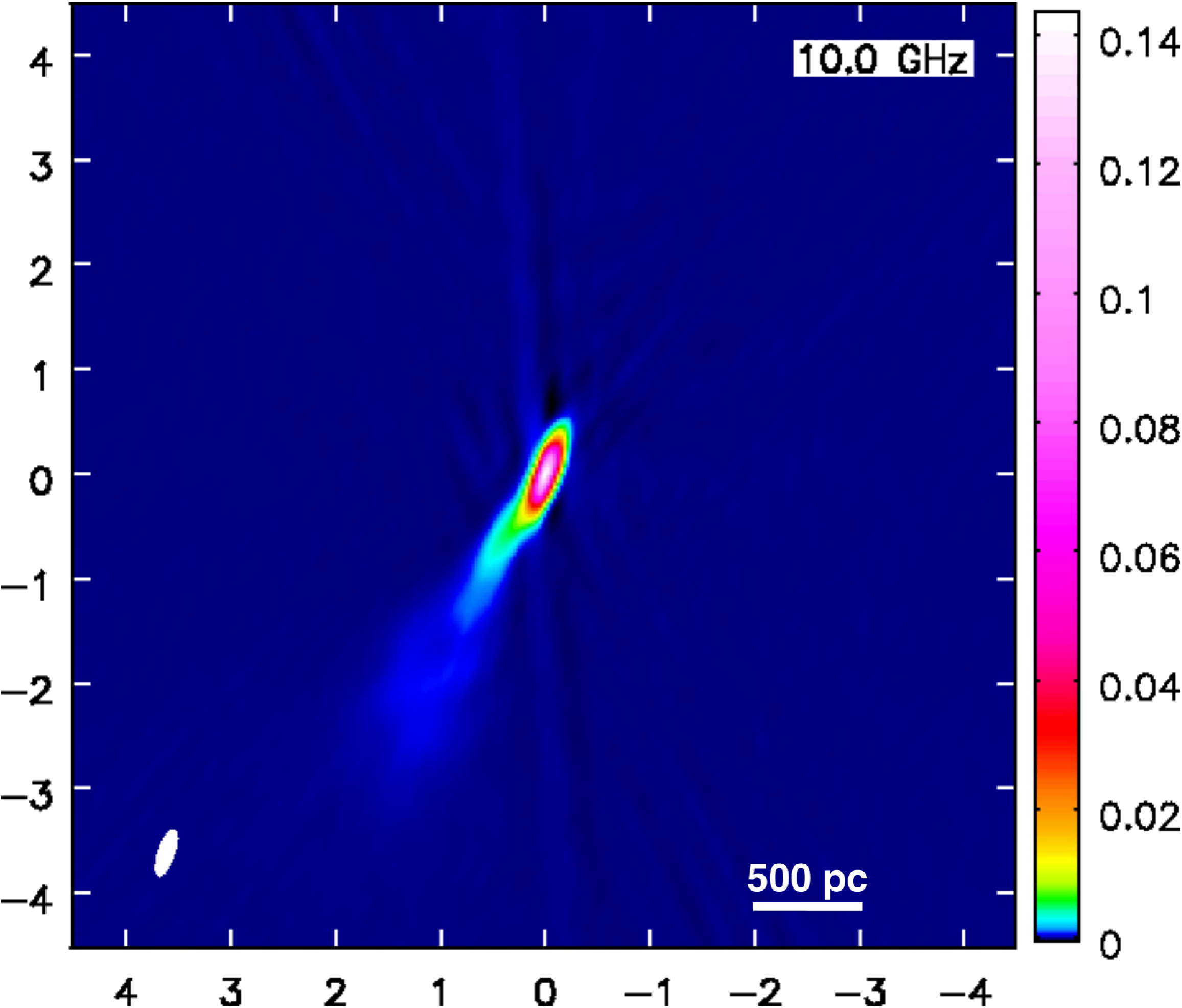}
\end{subfigure}
\hspace{3mm}
\begin{subfigure}[t]{.45\textwidth}
\centering
\caption{\textbf{NGC\,3100}}\label{fig:ngc3100_VLA}
\includegraphics[width=\linewidth]{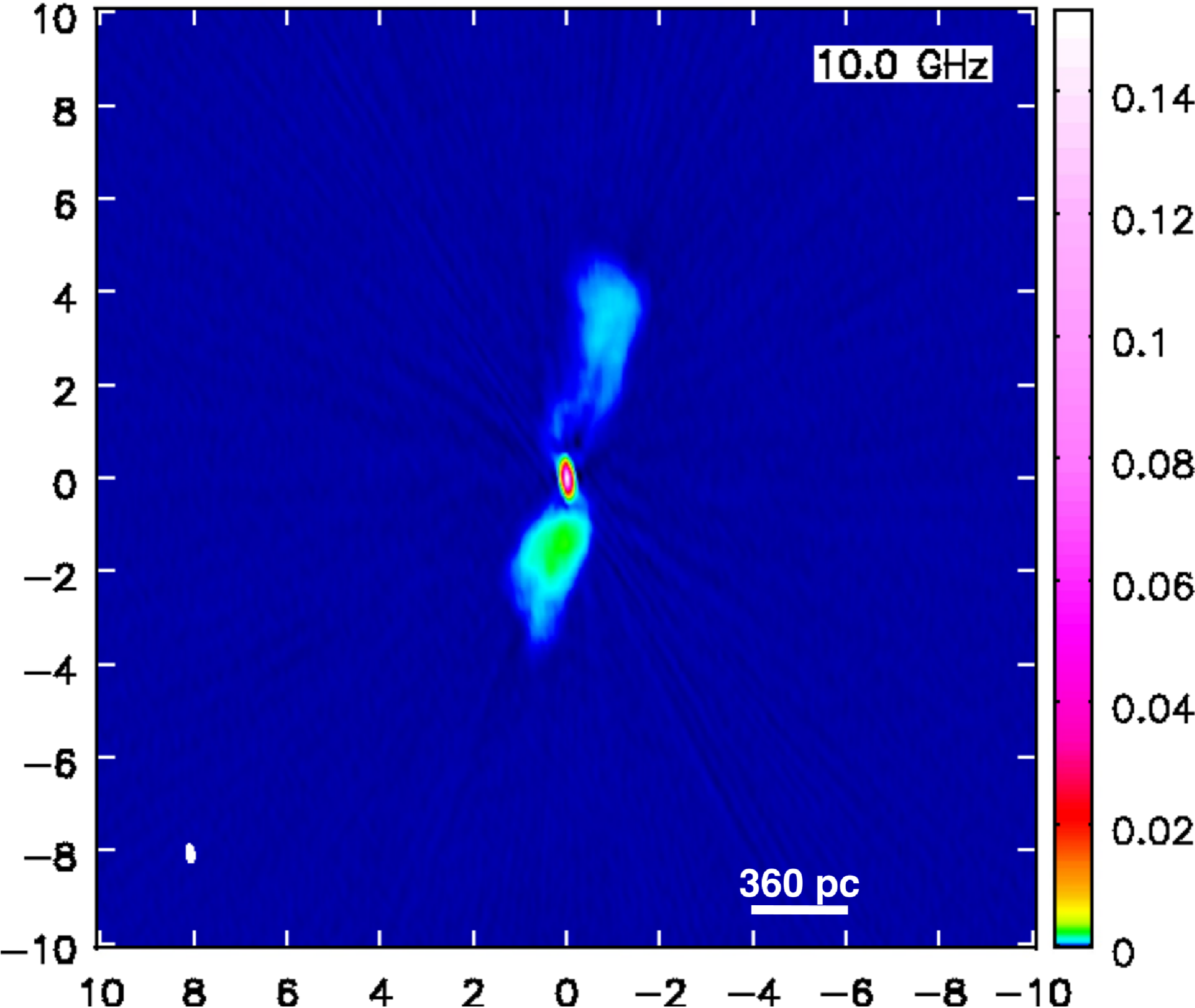}
\end{subfigure}

\medskip

\begin{subfigure}[t]{.48\textwidth}
\centering
\caption{\textbf{NGC\,3557}}\label{fig:ngc3557_VLA}
\includegraphics[width=\linewidth]{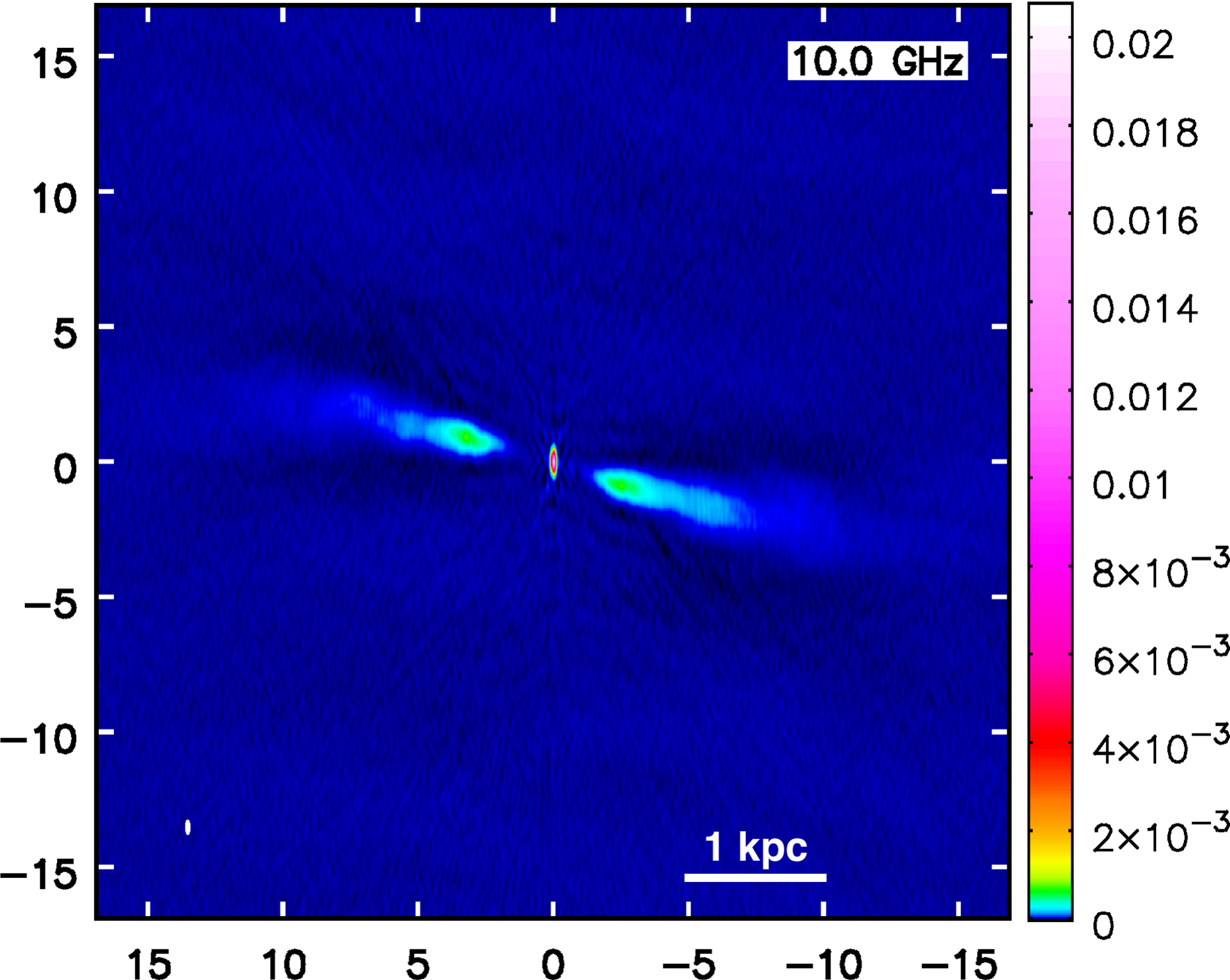}
\end{subfigure}

\medskip

\begin{subfigure}[t]{.47\textwidth}
\centering
 \caption{\textbf{IC\,4296}}\label{fig:ic4296_VLA}
\includegraphics[width=\linewidth]{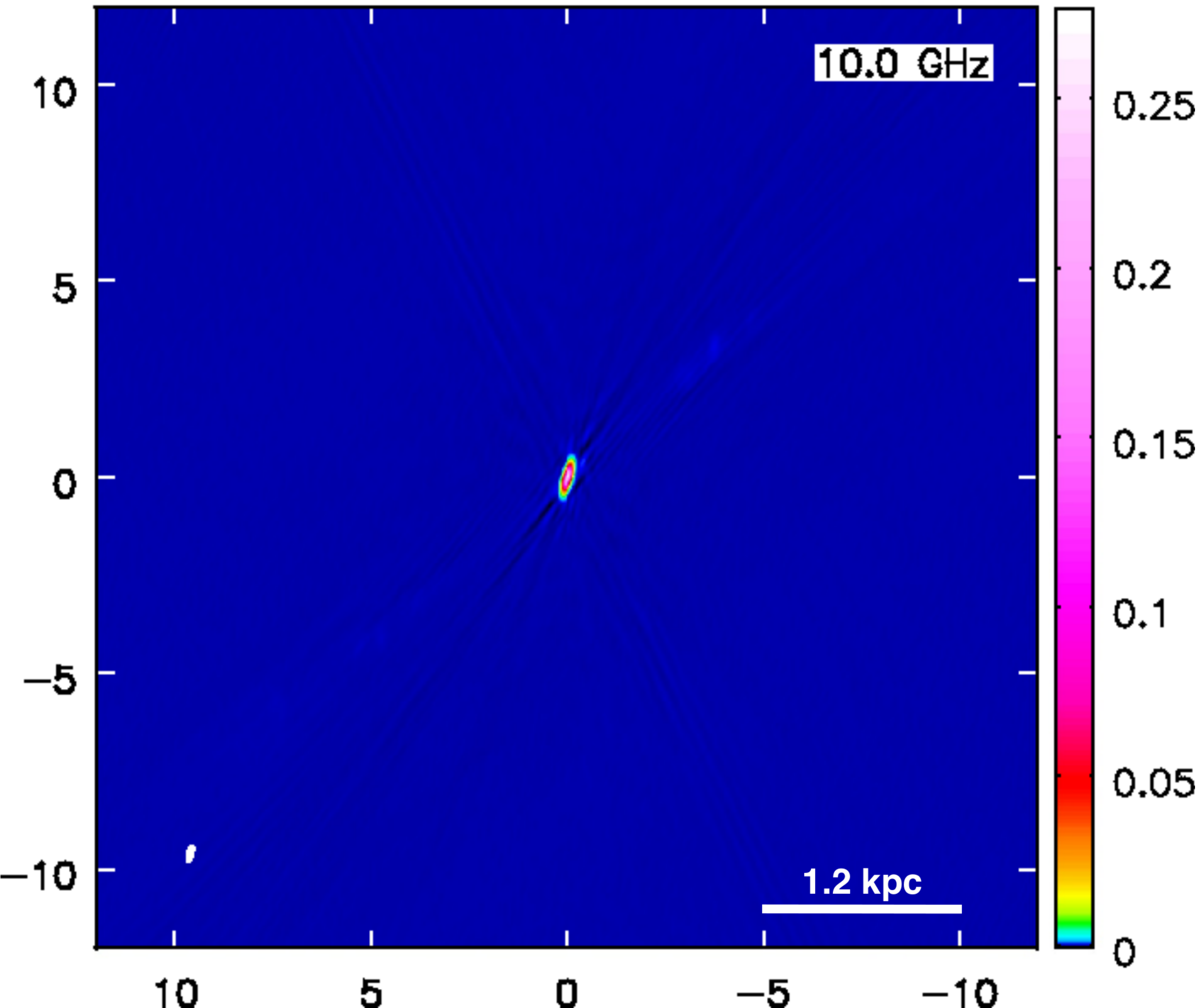}
\end{subfigure}
\hspace{3mm}
\begin{subfigure}[t]{.475\textwidth}
\centering
\caption{\textbf{NGC\,7075}}\label{fig:ngc7075_VLA}
\includegraphics[width=\linewidth]{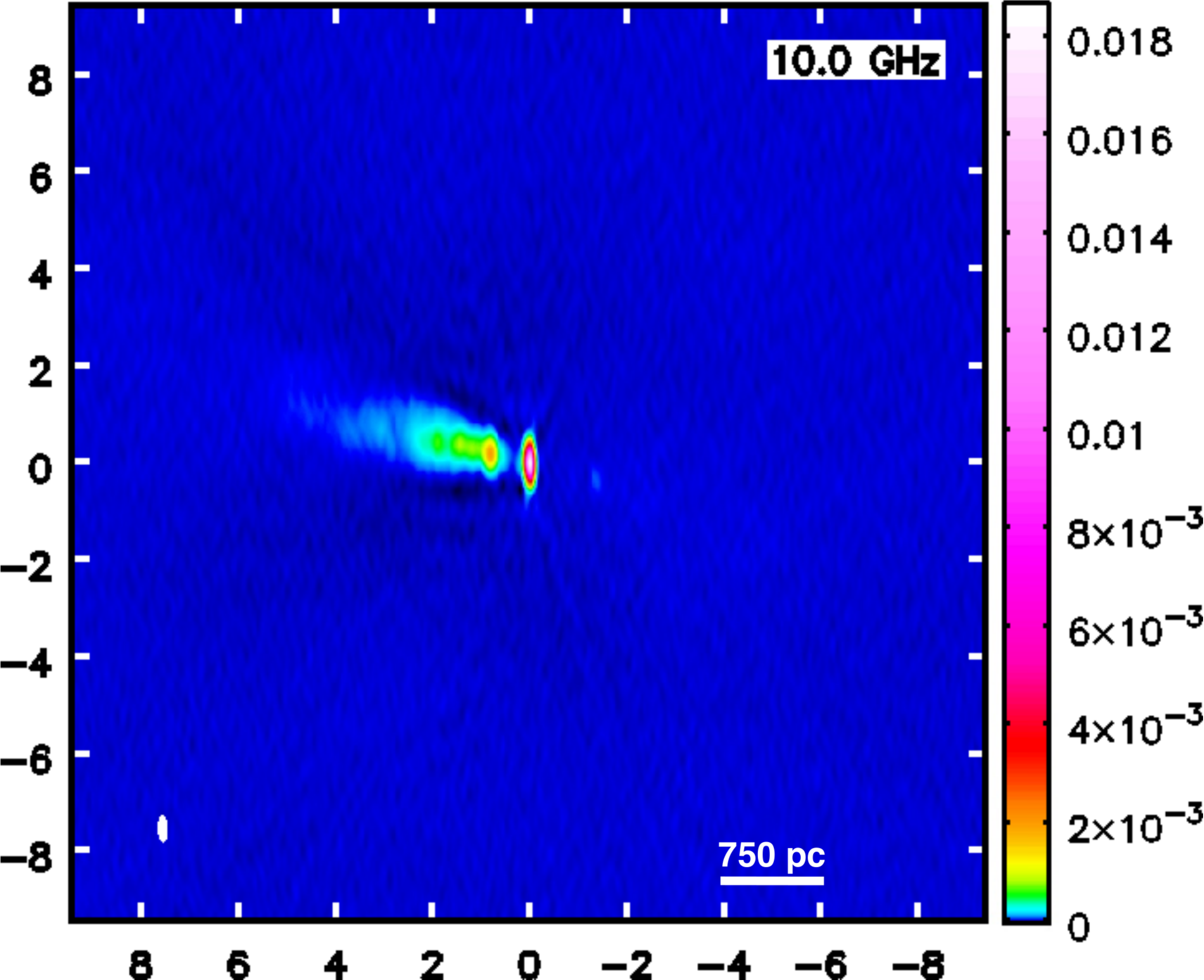}
\end{subfigure}
\caption[]{JVLA continuum maps at 10~GHz. The wedge on the right of each map shows the colour scale in Jy~beam$^{-1}$. Coordinates are given as relative positions with respect to the image phase centre in arcseconds; East is to the left and North to the top. The synthesised beam and scale bar are shown in the bottom-left and bottom-right corner, respectively, of each panel. \label{fig:VLAcont}}
\end{figure*}

Deconvolution and imaging were performed using the \textsc{CASA} \texttt{tclean} task in multi-frequency synthesis mode \citep{Rau11}. A second order Taylor series term was used (nterms~$=2$), to take into account the spectral behaviour of the sources. All of the continuum maps were made using Briggs weighting with robust $=0.0$, which gave a good trade-off between angular resolution and signal-to-noise ratio (S/N). Since the cores are detected at high S/N (between 50 and 262) in all of the targets, multiple cycles of phase-only and one cycle of amplitude and phase self-calibration were performed in all cases. Additional cycles of amplitude and phase self-calibration were performed for the brightest cores only. This allowed us to obtain root-mean square (rms) noise levels ranging from 4.8 to 14~$\mu$Jy~beam$^{-1}$ for synthesized beams of 0.37 -- 0.4\,arcsec full-width at half maximum  (FWHM; see Table~\ref{tab:JVLA obs}). The resulting maps are shown in Figure~\ref{fig:VLAcont}.

Two-dimensional Gaussian fits were performed within the regions covered by the continuum emission (identified as the regions where emission was detected at at least 5 times the rms noise level) to estimate the spatial extent of the observed components (when spatially resolved). Table~\ref{tab:VLA images} summarises the main parameters derived from the radio images. The quoted flux density errors include a 4\% contribution accounting for the flux calibration uncertainty of the JVLA data at 10~GHz \citep{Perley17}.

\begin{table*}
\centering
\caption{Radio properties of the targets as derived from the X-band JVLA continuum images.} 
\label{tab:VLA images}
\begin{tabular}{l c c c c c c c c c c c }
\hline
\multicolumn{1}{c}{ Target } & 
\multicolumn{1}{c}{ rms }&
\multicolumn{1}{c}{ S$_{\rm 10GHz}$ } &
\multicolumn{2}{c}{ Size FWHM } &
\multicolumn{1}{c}{ PA }\\
\multicolumn{1}{c}{  } & 
\multicolumn{1}{c}{ ($\mu$Jy~beam$^{-1}$)  } &
\multicolumn{1}{c}{ (mJy) } &
\multicolumn{1}{c}{ (arcsec$^{2}$) } &
\multicolumn{1}{c}{  (pc$^{2}$) } &
\multicolumn{1}{c}{ (deg) } \\
\multicolumn{1}{c}{  (1) } &
\multicolumn{1}{c}{  (2) } &
\multicolumn{1}{c}{  (3) } &
\multicolumn{1}{c}{  (4) } &
\multicolumn{1}{c}{  (5) } &
\multicolumn{1}{c}{  (6) } \\
\hline
IC\,1531 & 6.5 &    179$\pm$6 &     &   &    \\
core &     &     150$\pm$6 &  (0.13 $\times$ 0.06)  &    (70 $\times$ 30)  &  149$\pm$4  \\
 SE jet$^{*}$ &    &    28.8$\pm$1.1 &   $-$  &  $-$  &  $-$ \\
 \hline
NGC\,3100 & 5.1  &  244$\pm$6  &     &   &   \\
core &    &    156$\pm$6 &   (0.04 $\times$ 0.01)  &  (8.0 $\times$ 2.0)  &   17.3$\pm$6.4  \\
 N jet &   &  31.0$\pm$1.2  &   (2.3 $\times$ 0.9)  &  (440 $\times$ 170)  &  172$\pm$6  \\
S jet &   &  57.0$\pm$2.3  &   (1.4 $\times$ 0.8)  &  (290 $\times$ 130)  &  153$\pm$2  \\
\hline
NGC\,3557 &  4.8 &  57.8$\pm$1.6  &     &    &     \\
core$^{*}$ &    &    21.0$\pm$1.0 &   $-$  &  $-$ &  $-$  \\
E jet &    &  16.5$\pm$0.7 &   (1.8 $\times$ 0.5)  &  (380 $\times$ 110)  &  75$\pm$1  \\
W jet &    &  20.3$\pm$1.0  &    (4.3 $\times$ 0.6)  &  (915 $\times$ 130)  &  76$\pm$1   \\
\hline
IC\,4296 &  14  &    282$\pm$11  &     &    &   \\
core$^{*}$  &    &    282$\pm$11 &  $-$  &  $-$ &  $-$  \\
\hline
NGC\,7075 &  4.8  &     35.0$\pm$1.0  &     &    &    \\
core$^{*}$ &    &  19.0$\pm$0.8   &   $-$  &  $-$  &  $-$  \\
E jet &    &     15.8$\pm$0.6 &   (1.8 $\times$ 0.4)  &  (705 $\times$ 160)  &   77$\pm$1  \\
\hline
\end{tabular}
\parbox[t]{1\textwidth}{ \textit{Notes.} $-$ Columns: (1) Target name. (2) rms noise level measured in emission-free regions of the cleaned continuum maps in Figure~\ref{fig:VLAcont}. (3) 10~GHz continuum flux density. The total, core and jet flux densities are quoted separately, although extended regions are not well-defined in the 10~GHz continuum (Figure~\ref{fig:VLAcont}), thus their flux densities and associated uncertainties have to be considered as indicative only. The uncertainties are estimated as $\sqrt{{\rm rms}^{2} + (0.04 \times S_{\rm 10GHz})^{2}}$, and the second term dominates in all cases. (4) Size (FWHM) deconvolved from the synthesized beam. The sizes were estimated by performing 2D Gaussian fits to identifiable continuum components (see the text for details). (5) Spatial extent of each component corresponding to the angular size in column (4). (6) Position angle of the corresponding component, defined from North through East.\\
$^{*}$Unresolved component.}
\end{table*}

\section{Modelling of the radio jets and relative jet-disc orientations}\label{sec:method}
\subsection{Method}
Jets in FR\,I RGs are thought to be relativistic initially \citep[e.g.][]{Giovannini01,Hardcastle03} and then to decelerate rapidly on $\sim$kpc scales \citep[e.g.][]{Laing99}. Deceleration can be the result of injection of mass lost by stars within the jet volume \citep[e.g.][]{Komissarov94,Bowman96} or entrainment of the surrounding ISM \citep[e.g.][]{Baan80,Begelman82,Bicknell84,Bicknell86,DeYoung96,Rosen99,Rosen00}.
The kinematics of straight, twin radio jets in typical FR\,I radio galaxies have been extensively explored in a series of papers, coupling deep observations with accurate modelling \citep[see][for an overview]{Laing14}.
This analysis shows that the jets in FR\,I radio galaxies can be accurately modelled as intrinsically identical, antiparallel, axisymmetric, decelerating relativistic flows, and that the apparent surface brightness differences between them are dominated by relativistic aberration. The brighter (main) jet is then on the near side of the nucleus and the fainter counter-jet on the far side. In the vicinity of the core (i.e.\,within 1-2~kpc) three main features of the jet flow can be identified:
\begin{enumerate}
\item An inner, well-collimated region. In typical FR\,I radio galaxies (such as NGC\,383, the prototypical FR\,I LERG; \citealp[e.g.][]{Laing02}), the radio emission is weak in this region, particularly on the counter-jet side. 
\item A geometrical flaring region, where the jets first expand more rapidly and then re-collimate.  Within this region, the jets decelerate from $\approx0.9c$ to sub-relativistic speeds.
\item A flaring point within the geometrical flaring region. This is an abrupt increase in the surface brightness as a function of increasing distance from the core and marks the start of a region of high emissivity, associated with ongoing particle acceleration.
\end{enumerate}

Even for FR\,I jets that are too faint or poorly resolved to be modelled in detail, the systematic differences in apparent brightness between the main and counter-jet before the jets decelerate can be used to estimate inclinations to the line-of-sight. For intrinsically symmetrical, cylindrical, relativistic jets of constant velocity emitting isotropically in the fluid rest frame, the jet/counter-jet flux density ratio ($R$, also known as sidedness ratio) can be written as: 
\begin{eqnarray}\label{eq:sidedness_ratio}
R=\dfrac{I_{\rm jet}}{I_{\rm cjet}}=\left(\dfrac{1+\beta \cos\theta_{\rm jet}}{1-\beta \cos\theta_{\rm jet}}\right) ^{2-\alpha}
\end{eqnarray}
\citep{BK79}, where $\theta_{\rm jet}~(0\leq \theta_{\rm jet} \leq \pi/2)$ is the angle to the line-of-sight of the approaching jet (i.e.\,the jet inclination angle), $\beta=v/c$ (where $v$ is the flow velocity) and $\alpha$ is the jet spectral index ($S\propto\nu^{\alpha}$).

\begin{figure}
\centering
\includegraphics[scale=0.39]{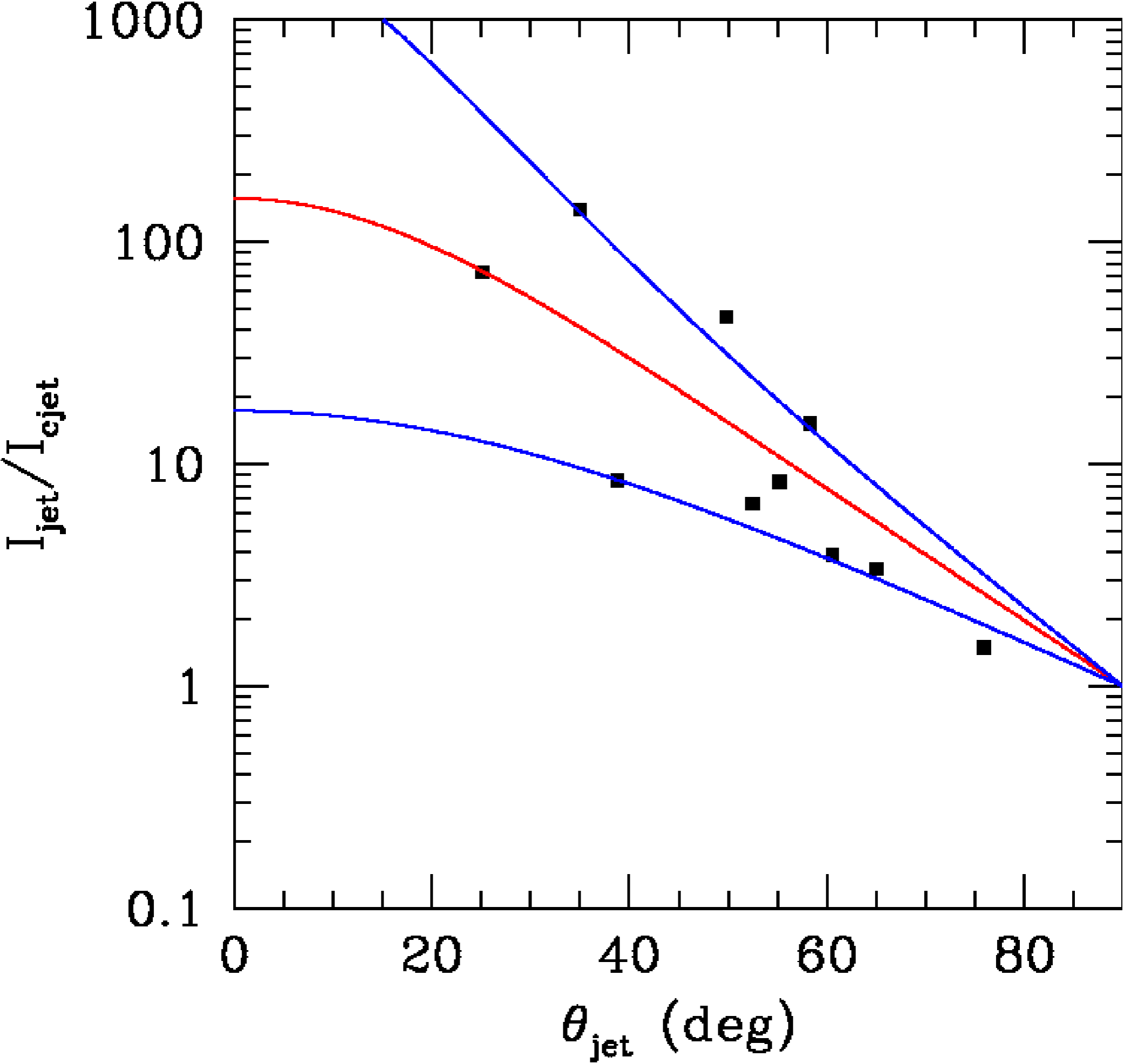}
\caption[]{\small{Jet/counter-jet flux density ratio versus jet inclination angle for the 10 RGs studied in \citet{Laing14}. The spread in sideness ratio is dominated by intrinsic differences between objects
rather than measurement errors, which are therefore not shown. The sidedness ratios were measured as in the present paper from the images listed in Table~2 of \citet{Laing14}. The solid lines represent single-velocity models (Equation~\ref{eq:jet_inclination}) for $\beta=0.75$ (red, best fit), $\beta = 0.5$ and $\beta = 0.9$ (both blue).}}\label{fig:ratio_vs_inclination}
\end{figure}

Equation~\ref{eq:sidedness_ratio} does not allow $\beta$ and $\theta_{\rm jet}$ to be determined independently. However, the detailed modelling carried out by \citet{Laing14}, which also uses constraints from linear polarization, enables velocities and inclinations to be decoupled and accurate values of $\theta_{\rm jet}$ have been derived for ten FR\,I RGs. The inferred velocity fields show longitudinal and transverse variations, but the dispersion in the jet velocity just downstream of the brightness flaring point is sufficiently small that we can fit the empirical relation between $R$ and $\theta_{\rm jet}$ using a simple constant-velocity approximation. This in turn enables an empirical calibration of the relation, which we assume to be valid for FR\,I jets in general.  Figure~\ref{fig:ratio_vs_inclination} shows a version of the R$-\theta_{\rm jet}$ relation adapted for the present paper by measuring the jet sidedness ratios of the ten RGs studied in \citet{Laing14} following the same method adopted later for our sources (see Section~\ref{sec:sidedness_ratio}). The solid lines in Figure~\ref{fig:ratio_vs_inclination} are single-velocity models. The model we adopt (red solid line) has $\beta = 0.75$, determined from an unweighted least-squares fit of Equation~\ref{eq:sidedness_ratio} to the data in Figure~\ref{fig:ratio_vs_inclination}. We take $\alpha=-0.6$, which is the mean spectral index typically measured between 1.4 and 4.9~GHz in FR\,I jet bases \citep{LB13}. ALMA observations of the jets confirm that  there is little or no spectral steepening up to 230~GHz (Paper I) and therefore that this value should apply at 10~GHz.

Once the value of $\beta$ is fixed, a reasonable estimate of the inclination of the jet to the line of sight can be obtained by simply inverting Equation~\ref{eq:sidedness_ratio}:
\begin{eqnarray}\label{eq:jet_inclination}
\theta_{\rm jet} &=& \arccos \left( \frac{1}{\beta} \frac{R^{\frac{1}{2 - \alpha}} - 1 }{R^{\frac{1}{2 - \alpha}} +1 }  \right).
\end{eqnarray}

\begin{figure*}
\centering
\begin{subfigure}[t]{.365\textwidth}
\centering
\caption{\textbf{NGC\,3557}}\label{fig:ngc3557_boxes}
\includegraphics[width=\linewidth]{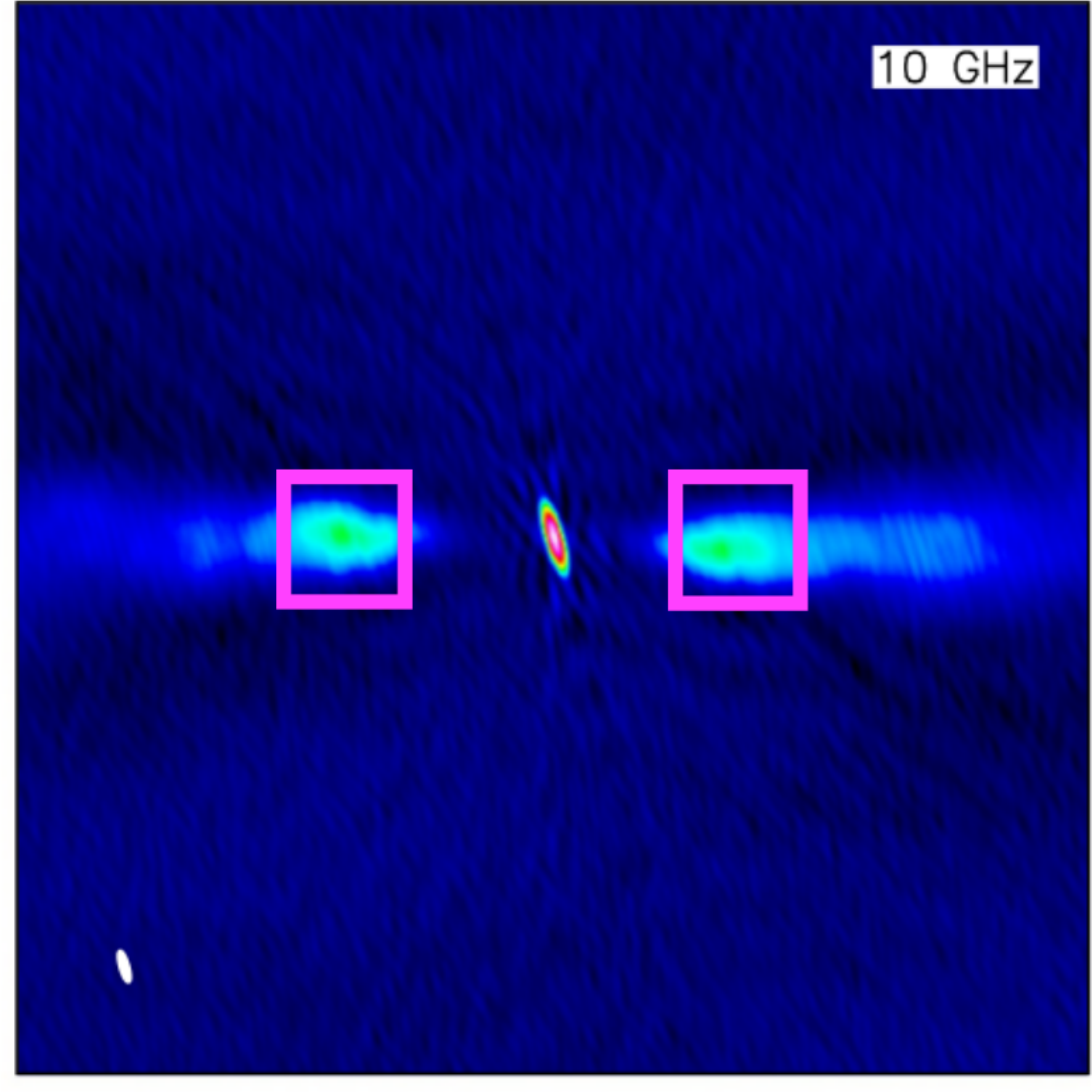}
\end{subfigure}
\begin{subfigure}[t]{.36\textwidth}
\centering
\caption{\textbf{NGC\,7075}}\label{fig:ngc7075_boxes}
\includegraphics[width=\linewidth]{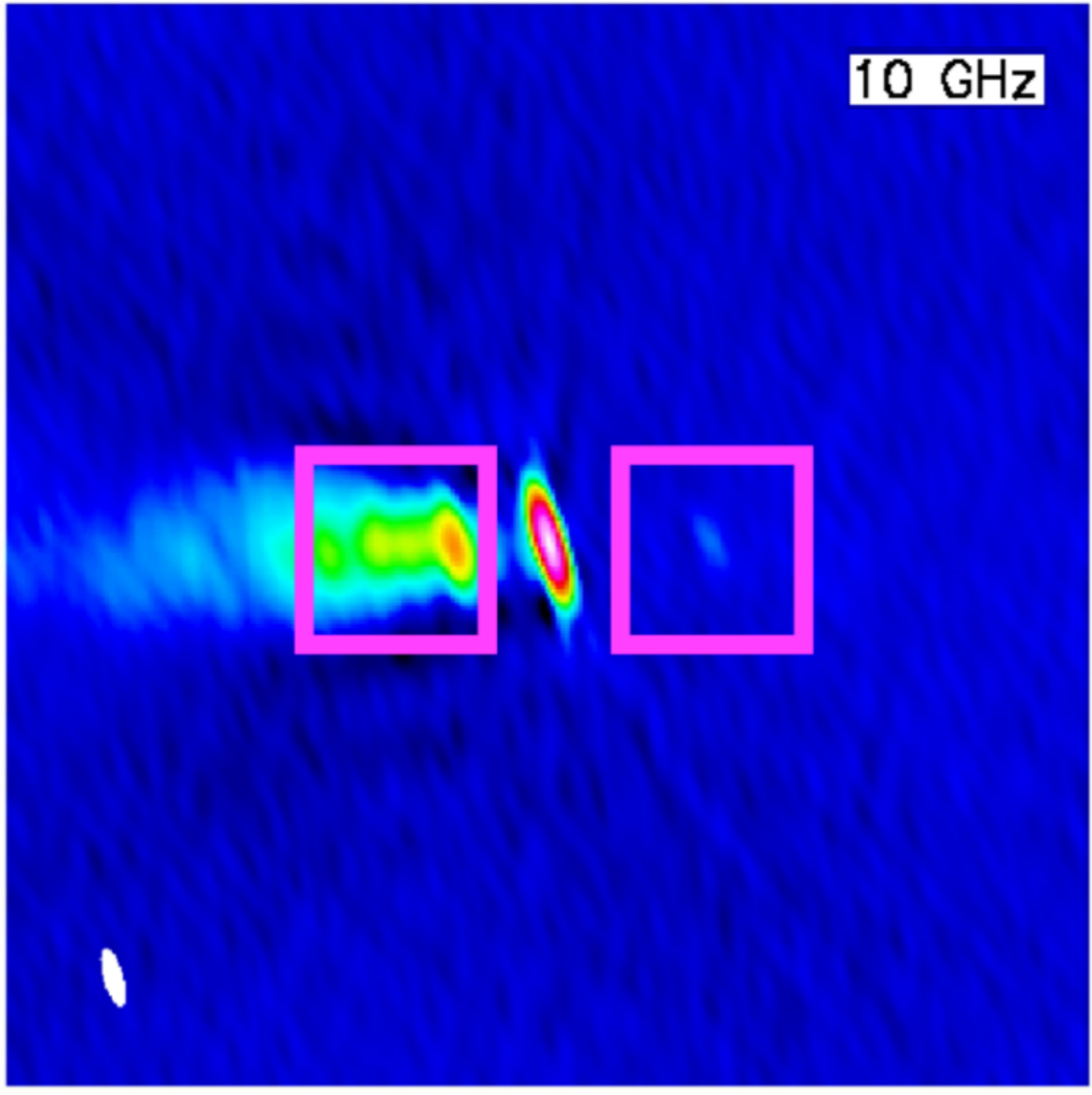}
\end{subfigure}

\medskip

\begin{subfigure}[t]{.365\textwidth}
\centering
\caption{\textbf{IC\,1531}}\label{fig:ic1531_boxes}
\includegraphics[width=\linewidth]{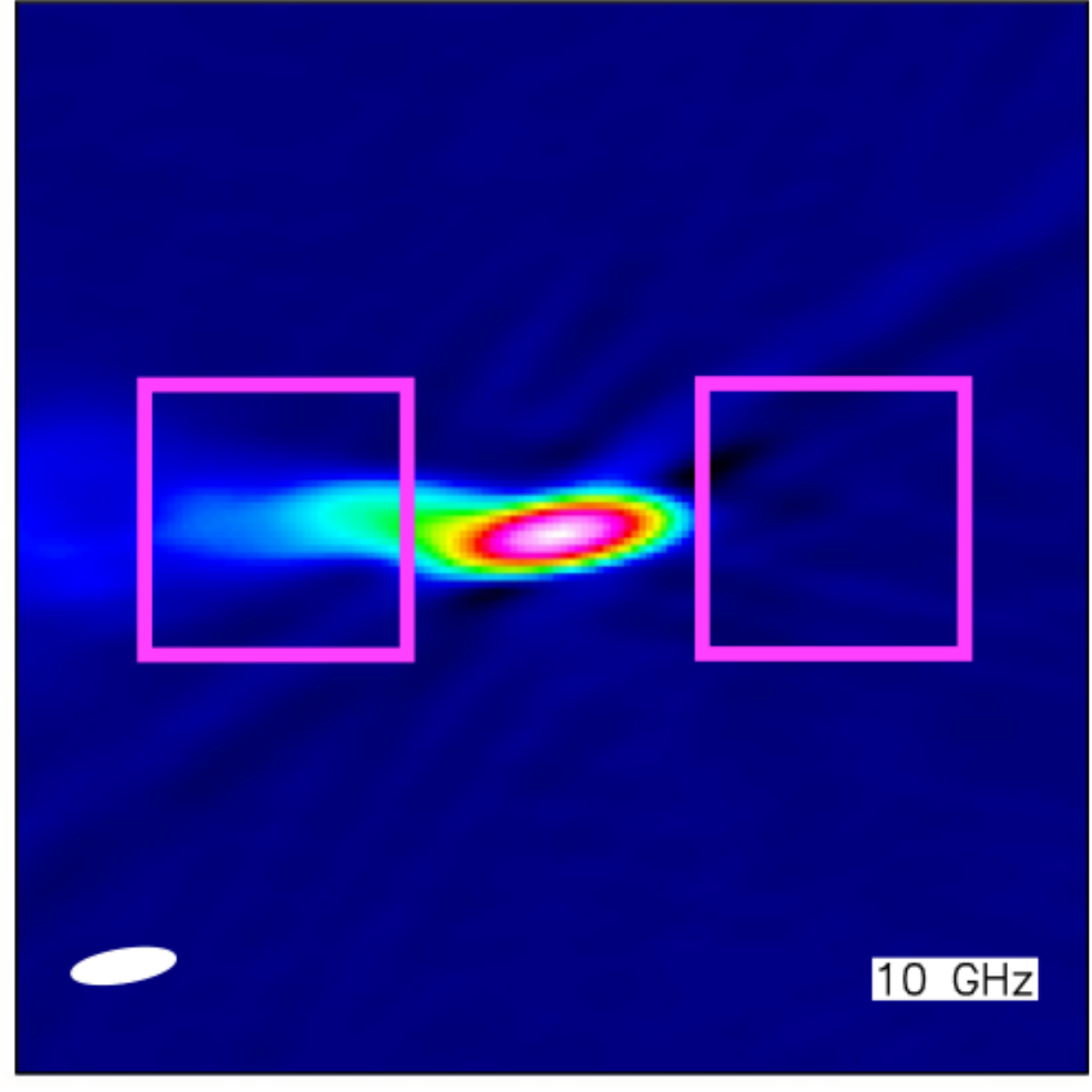}
\end{subfigure}
\begin{subfigure}[t]{.36\textwidth}
\centering
 \caption{\textbf{IC\,4296}}\label{fig:ic4296_boxes}
\includegraphics[width=\linewidth]{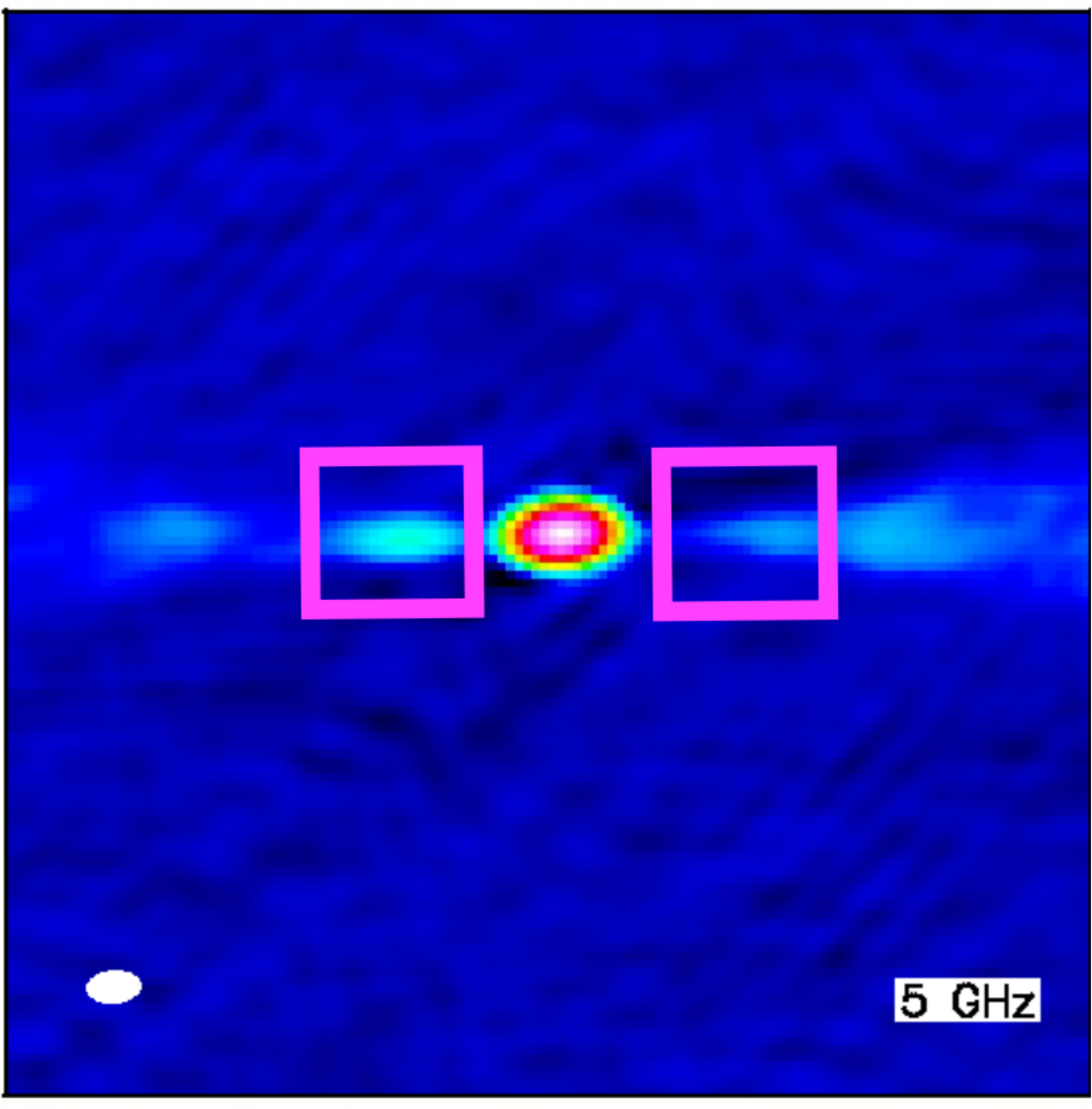}
\end{subfigure}
\caption[]{Radio continuum maps of the four southern sample sources for which we have estimated the jet inclination to the line of sight. Panels \textbf{a}, \textbf{b} and \textbf{c} show rotated versions of the 10~GHz JVLA maps of NGC\,3557, NGC\,7075 and IC\,1531, respectively. Panel \textbf{d} shows a rotated version of the archival VLA image of IC\,4296 at 5~GHz presented in Paper I (see the text for details). The boxes used to measure the jet and counter-jet flux densities are overlaid in magenta. The beam size and the reference frequency are indicated on each panel. \label{fig:VLAboxes}}
\end{figure*}

\subsection{Jet sidedness ratios and inclination estimates}\label{sec:sidedness_ratio}

To estimate $\theta_{\rm jet}$ for each object using Equation~\ref{eq:jet_inclination}, we derive $R$ by measuring the jet/counter-jet flux density ratio just downstream of the brightness flaring point. The reasons for this choice are that this location is easily identifiable in high-resolution images of FR\,I jets \citep{Laing99} and that flow velocities are high, with a fairly small dispersion between sources \citep{Laing14}. The empirical method adopted to estimate $\theta_{\rm jet}$ is as follows:
\begin{enumerate}
\item set boxes covering the brightness flaring region symmetrically on both sides of the nucleus: the inner boundary is set at the flaring point and the box extends away from the core for three synthesized beams;
\item measure the total flux density in each box using the \textsc{CASA} task \texttt{imstat}; 
\item divide the flux density of the (brighter) jet by that of the counter (fainter) jet to obtain $R$, and then use Equation~\ref{eq:jet_inclination} with $\beta = 0.75$ to estimate the jet inclination angle, $\theta_{\rm jet}$.
\end{enumerate}

The locations of the used boxes are shown on rotated maps of the inner jet regions in  Figure~\ref{fig:VLAboxes}.

Measured sidedness ratios ($R$) and derived jet inclination angles ($\theta_{\rm jet}$) are listed in Table~\ref{tab:alignments}.  We consider four sources of error in our estimates of $\theta_{\rm jet}$:
\begin{enumerate}
\item Dispersion in the $R - \theta_{\rm jet}$  relation is modelled by calculating $\theta_{\rm jet}$ for $\beta = 0.5$ and $\beta = 0.9$, which bound the distributions in Figure~\ref{fig:ratio_vs_inclination} at approximately 68\% confidence.
\item Errors in flux-density measurements are estimated from the distribution of values for boxes of the same size distributed over many source-free locations on the image and propagated into the estimates of $R$. 
\item The effect of intrinsic asymmetries in flux density is modelled as a 20\% error on $R$, based on the analysis of \citet[][Section~7.2]{Laing14}.
\item Measurement errors and intrinsic dispersion in spectral index $\alpha$ have a minor effect on the derived value of $\theta_{\rm jet}$ ($\la 1^{\circ}$) and are not included.
\end{enumerate}
Contributions (i) $-$ (iii) are propagated through Equation~\ref{eq:jet_inclination} and summed in quadrature to give the error
estimates on $\theta_{\rm jet}$ in Table~\ref{tab:alignments}. Item (i) dominates for small $\theta_{\rm jet}$; (ii) and (iii) at larger inclinations. 

The method outlined above was applied to NGC\,3557 and NGC\,7075 using our new 10-GHz images. In these cases, the inner FR\,I jet structure is well resolved, although in NGC\,7075 emission from the counter-jet is only marginally visible at the flaring point (Figs.~\ref{fig:ngc3557_boxes} and \ref{fig:ngc7075_boxes}). In 
IC\,4296, emission from the brightest knots of the North-West jet is only barely visible in our 10-GHz observations (Fig.~\ref{fig:ic4296_VLA}) and emission from the South-East jet is undetected. We therefore used deep high-resolution archival VLA data at 5~GHz (whose re-imaged version is presented in Paper I and Figure~\ref{fig:ic4296_boxes}) to estimate the jet sidedness ratio.

In IC\,1531, the inner jet structure is unresolved at the spatial resolution of our 10~GHz observations ($\approx200$~pc; see Table~\ref{tab:JVLA obs}), the flaring point is not clearly identifiable and emission from the counter-jet is not securely detected, qualitatively consistent with its blazar classification \citep{Bassi18} and a small value of  $\theta_{\rm jet}$.  In this case, we placed the two boxes as close to the core as possible (Fig.~\ref{fig:ic1531_boxes}). We then derived a $3\sigma$ upper limit to the flux density of the counter-jet from the distribution for source-free regions of the image. The lower limit on $R$ corresponds to an upper limit $\theta_{\rm jet} \lesssim 19^\circ$.  \citet[][]{Bassi18} presented a detailed multi-wavelength study of IC\,1531, and estimated the jet inclination angle to be $\theta_{\rm jet} = 15^\circ \pm 5^\circ$, consistent with this limit. We adopt their value.

As anticipated in Section~\ref{sec:obs}, NGC\,3100 is a peculiar case which does not show the normal FR\,I jet geometry on sub-kpc scales (see Figure~\ref{fig:ngc3100_VLA}).  We therefore cannot estimate the jet inclination under the simple assumption of straight, intrinsically symmetrical  relativistic twin jets. This special case will be discussed separately in Section~\ref{sec:ngc3100_VLA_discuss}.

To improve the statistics, we have expanded the sample by including NGC\,383 and NGC\,3665, two other FR\,I LERGs for which all of the data needed to perform our analysis are available in the literature. The jet inclination for NGC\,383 is derived from the detailed modelling carried out by \citet{Laing02,Laing14}. For NGC\,3665, we take the sidedness ratio from \citet{Laing99}, which was derived in an identical way to those in this paper (albeit at 1.4\,GHz), and then use Equation~\ref{eq:jet_inclination} to derive $\theta_{\rm jet}$. The jet parameters of these two additional sources are also listed in Table~\ref{tab:alignments}.

\subsection{Relative jet-CO disc orientation in 3D}\label{sec:jet_disc_inclination}
Once an estimate of the line-of-sight jet inclination is obtained, we can attempt to constrain the {\em intrinsic} orientation of the jet axis relative to the rotation axis of the CO disc ($\theta_{\rm dj}$).

To do this, for each object we make use of:
\begin{enumerate}
\item the jet inclination angle ($\theta_{\rm jet}$), derived as described in Section~\ref{sec:sidedness_ratio};
\item the CO disc inclination angle ($\theta_{\rm disc}$) obtained from the 3D modelling of the discs reported in Paper II (southern sample),  \citealt{North19} (NGC\,383) and \citealt{Onishi17} (NGC\,3665);
\item  the CO disc rotation axis position angle (PA$_{\rm rot}$), defined as the position angle projected onto the plane
of the sky of a vector along the CO disc rotation axis pointing towards the
observer (i.e.\,on the near side of the disc), measured anticlockwise from North and ranged $\pm180^{\circ}$. This is derived from the best-fit kinematic position angles (PA$_{\rm kin}$) listed in Table~\ref{tab:Southern Sample};
\item the position angle of the approaching radio jet (PA$_{\rm jet}$), i.e.\, the position angle of the approaching jet axis as it appears projected onto the plane of the sky, measured anticlockwise from North and ranged $\pm180^{\circ}$ (taken from the radio images used to estimate sidedness ratio as described in Section~\ref{sec:sidedness_ratio});

\item the {\em projected} relative jet-disc orientation ($\Delta$) derived from the difference between the angles (iii) and (iv), i.e.\,$\Delta = |PA_{\rm rot} - PA_{\rm jet}|$, ranged $0^\circ - 180^\circ$. 
\end{enumerate}

The intrinsic angle between the jet and the disc rotation axis,
$\theta_{\rm dj}$, can be determined from $\theta_{\rm jet}$, $\theta_{\rm disc}$ and $\Delta$ using the relation:

\begin{eqnarray}\label{eq:inclination_angle}  
  \cos\theta_{\rm dj} = |\cos\Delta \sin \theta_{\rm disc}\sin \theta_{\rm jet} + \cos \theta_{\rm disc}\cos \theta_{\rm jet}|.
\end{eqnarray}

The mathematical steps from which this relation is derived are reported in Appendix~\ref{sec:Appendix}, where the model parameters and coordinate systems we use to describe the 3D jet-disc geometry are also sketched  (Figure~\ref{fig:jet_disc_geometry}).

We note that we cannot distinguish between the near and far sides of the disc using CO observations alone (at least at the spatial resolution of the data used in the present work). In our cases, this ambiguity can be removed only if additional information is available. For instance, if we observe dust associated with the CO disc, we can infer that the near side of the disc is the one  where the dust absorption against the underlying stellar light is stronger. If no such information is available, there is a $180^{\circ}$ ambiguity in the value of PA$_{\rm rot}$ and therefore in $\Delta$, implying that there are two possible values of $\theta_{\rm dj}$. For NGC\,3557, IC\,4296, NGC\,383 and NGC\,3665 the available archival Hubble Space Telescope (HST) optical imaging allows us to identify the side of the CO disc pointing towards the observer unequivocally (Paper~I; \citealt{Onishi17,North19}) and we can derive unique values of $\theta_{\rm dj}$. For IC\,1531 and NGC\,7075, no such information is available and we report both possible values.

\begin{figure}
\centering
\includegraphics[scale=0.31]{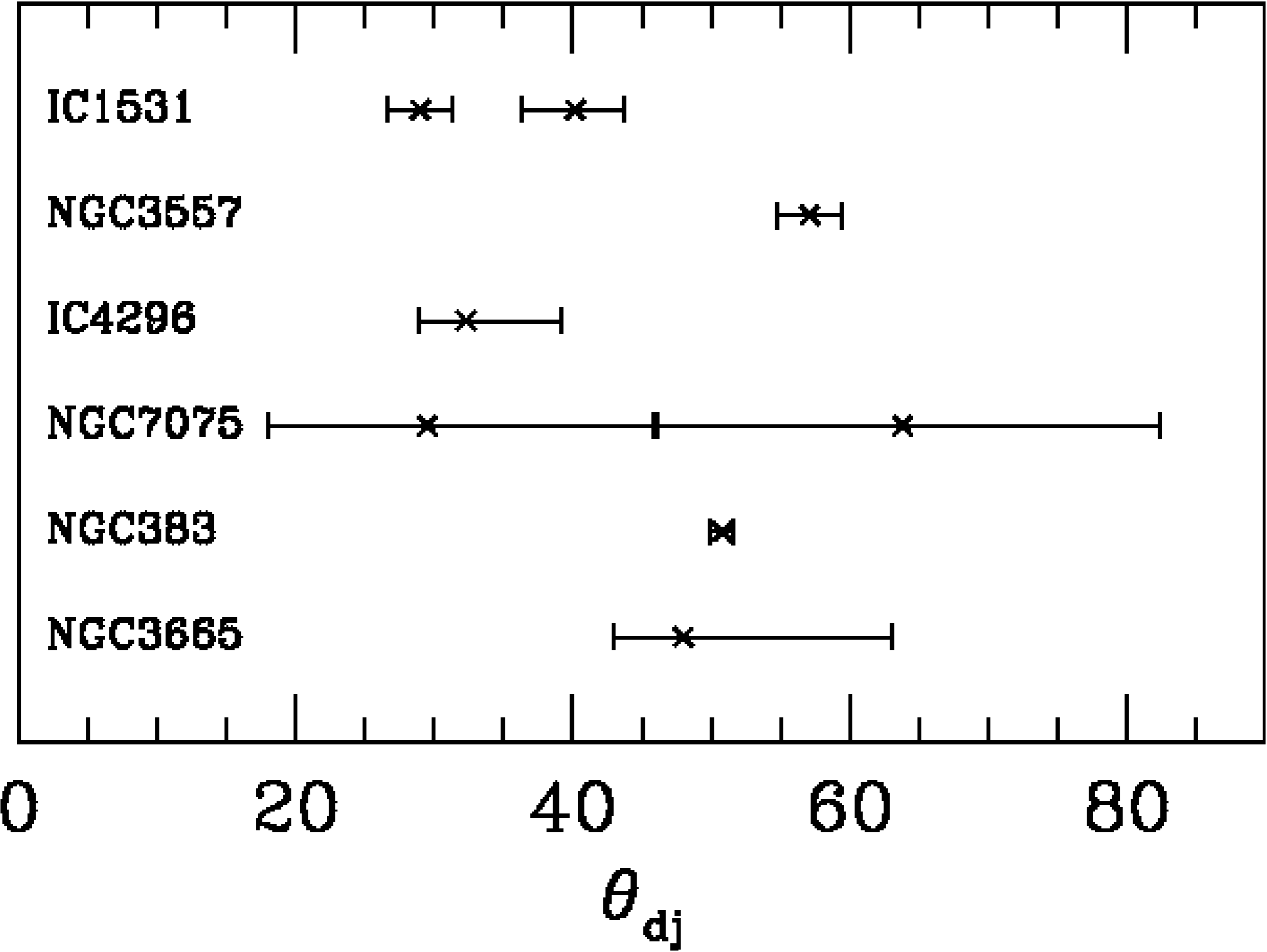}
\caption[]{\small{Distribution of intrinsic jet-disc orientation angles, $\theta_{\rm dj}$ from Table~\ref{tab:alignments}. For IC\,1531 and NGC\,7075, the disc orientation is ambiguous and both possible values are plotted.}}\label{fig:thetadj}
\end{figure}

Values of $\theta_{\rm disc}$, PA$_{\rm rot}$, PA$_{\rm jet}$ and $\Delta$ are listed in Table~\ref{tab:alignments}. The final intrinsic jet-disc orientation angles, ($\theta_{\rm dj}$),  are also listed in Table~\ref{tab:alignments} and plotted in Figure~\ref{fig:thetadj}. Errors on $\theta_{\rm dj}$ are determined by propagating those for $\theta_{\rm disc}$, $\theta_{\rm jet}$ and
$\Delta$ through Equation~\ref{eq:inclination_angle} and summing in quadrature.

\begin{table*}
\centering
\begin{scriptsize}
\caption{Jet and CO disc parameters and relative orientation angles.}
\label{tab:alignments}
\begin{tabular}{l r c c c c c c c}
\hline
\multicolumn{1}{c}{ Target } &
\multicolumn{1}{c}{ $R$ } & 
\multicolumn{1}{c}{ $\theta_{\rm jet}$ } &
\multicolumn{1}{c}{ $\theta_{\rm disc}$ } & 
\multicolumn{1}{c}{PA$_{\rm jet}$} &
\multicolumn{1}{c}{PA$_{\rm rot}$} &
\multicolumn{1}{c}{ $\Delta$ } &
\multicolumn{1}{c}{ $\theta_{\rm dj}$ } &
\multicolumn{1}{c}{ References }\\ 
\multicolumn{1}{c}{  } &    
\multicolumn{1}{c}{  } &
\multicolumn{1}{c}{  (deg)} &   
\multicolumn{1}{c}{  (deg)} &   
\multicolumn{1}{c}{  (deg)} &
\multicolumn{1}{c}{  (deg)} &
\multicolumn{1}{c}{  (deg)} &
\multicolumn{1}{c}{  (deg)} &   
\multicolumn{1}{c}{   } \\
\multicolumn{1}{c}{   (1) } &   
\multicolumn{1}{c}{   (2) } &
\multicolumn{1}{c}{   (3) } &
\multicolumn{1}{c}{   (4)} &
\multicolumn{1}{c}{   (5)} &
\multicolumn{1}{c}{   (6)} &
\multicolumn{1}{c}{   (7)} &
\multicolumn{1}{c}{   (8)} &
\multicolumn{1}{c}{   (9)} \\
\hline
\vspace{1.5mm}
IC\,1531$^1$ &  $\geq$96\;\;\;\; & $15.0\pm 5.0$ & $32.0\pm 2.5$ & $152.0\pm 2.0$ &$86.0\pm 1.0$ ($-94.0\pm 1.0$) &$66.0 \pm 1.7$ ($114.0 \pm 1.7$)& $29.0\pm 2.3$ ($40.2^{+3.5}_{-3.9}$)& a, b, c\\
\vspace{1.5mm}
NGC\,3557  &   1.0$\pm0.1$   & $90.0^{+0.0}_{-3.0}$  & $56.0\pm 1.0$ & $73.0\pm 1.0$ ($-108.0\pm 1.0$) & $-59.0\pm 1.0$ & $ 49.0\pm 1.4$ ($131.0\pm 1.4$)  &  $57.1^{+2.3}_{-2.4}$ & a, b, c \\ 
\vspace{1.5mm}
 IC\,4296$^2$  &  $1.8\pm0.1$   & 81.4$^{+2.7}_{-5.2}$  &  $68.0\pm 1.5$ &   $-50.5\pm 1.0$&$140.0\pm 3.0$&$169.5 \pm 3.2$  & $32.3^{+5.3}_{-3.1}$ & b, c \\  
 NGC\,7075 &   $99.0\pm 9.9$  &  $19.2^{+19.5}_{-19.2}$  & $46.0\pm 1.5$ &$77.0\pm1.8$ & $52.0\pm 4.0$ ($-128.0\pm 4.0$)& $25.0 \pm 4.4$ ($155.0\pm 4.4$)&  $29.5^{+16.6}_{-11.4}$  ($63.8^{+18.6}_{-17.9}$)& a, b, c \\ 
\hline
\multicolumn{9}{c}{\small{\textbf{Literature data}}}\\
\hline
\vspace{1.5mm}
3C31$^3$ &  6.6$\pm 0.3$ & $52.5^{+0.5}_{-0.9}$  & $37.6\pm 1.7$ &$-19.7\pm 1.0$&$52.2$& $71.9\pm 1.0$ & $50.8^{+0.7}_{-0.9}$  & d, e, f \\
NGC\,3665 &   6.1$\pm 0.6$ & $63.5^{+5.6}_{-15.6}$  & $69.9$ & $-52.2\pm 2$ & $116.0$ & $168.2\pm 2.0$ & $48.0^{+15.1}_{-5.1}$  & g, h \\ 
\hline 
\end{tabular}
\end{scriptsize}
\parbox[t]{1\textwidth}{ \textit{Notes.} $-$ Columns: (1) Target name. (2) Jet sidedness ratio measured as described in Section~\ref{sec:sidedness_ratio}. (3) Jet inclination, i.e.\,angle between the approaching jet and the line of sight, ranged in $0^{\circ} - 90^\circ$. (4) CO disc inclination, i.e.\,angle between the line of sight and the axis normal to the disc plane on the near side of the disc, ranged in $0^{\circ} - 90^\circ$. (5) Position angle of the approaching jet axis as it appears projected onto the plane of the sky, ranged $\pm180^{\circ}$. Two values differing by $180^\circ$ are given for NGC\,3557, where the jets are in the plane of the sky (i.e.\,the jet and counter-jet are indistinguishable). (6) Position angle of the CO disc rotation axis, i.e.\,position angle projected on the plane
of the sky of a vector along the rotation axis pointing towards the
observer, measured anticlockwise from North and ranged $\pm180^{\circ}$. This is derived from the best-fit kinematic position angles (PA$_{\rm kin}$) listed in Table~\ref{tab:Southern Sample}. If the near and far sides of the CO disc cannot be distinguished, there is an $180^{\circ}$ ambiguity in PA$_{\rm rot}$ and both values are quoted (see Section~\ref{sec:method}  and Appendix~\ref{sec:Appendix} for details). (7) Projected relative orientation between the jet and the rotation axis of the CO disc, derived from the values in columns (5) and (6), i.e.\,$\Delta = | PA_{\rm rot} - PA_{\rm jet}|$, ranged 0$^{\circ}$ to $180^{\circ}$. Two values are quoted in case of ambiguity in either the jet or the disc position angle. (8) Intrinsic relative orientation angle between the jet and the CO disc rotation axis. Two values are quoted for IC\,1531 and NGC\,7075, due to the ambiguity in the near/far sides of the disc. The two values of $\Delta$ for NGC\,3557 give identical results for $\theta_{\rm dj}$. (9) References for the jet and CO disc parameters used to derive the relative orientation angles listed in column 8. (a): This work, (b): \citet{Ruffa19a}, (c): \citet{Ruffa19b}, (d): \citet{Laing14}, (e): \citet{North19},  (f):  \citet{Laing02}, (g):\citet{Laing99}, (h): \citet{Onishi17}.\\

$^1$ The jet inclination is taken from \citet{Bassi18}, which is consistent with our lower limit on $R$.

$^2$ Sidedness ratio estimated from archival radio data at 5~GHz (Paper~I; see text for details).

$^3$ $\theta_{\rm jet}$ not obtained from sidedness ratio, but from full modelling of the radio jets (see \citealp[][]{Laing14}).}
\end{table*}

\section{Results and discussion}\label{sec:results}
\subsection{Intrinsic jet-disc relative orientations}\label{sec:thetadj}
Simulations of jet formation in low accretion-rate systems like LERGs show that jets are launched along the spin axis of the central black hole, which is in turn aligned with the inner accretion disc  \citep[e.g.][]{McKinney12}. In a simple axisymmetric system, we might expect the accretion disc to be co-axial with the (sub-)kpc scale molecular disc and therefore that the jet and disc rotation axes would be accurately aligned. Earlier work on projected misalignments (summarized in Section~\ref{sec:intro}) has already shown that this picture is oversimplified. Our new results add estimates of misalignments in three dimensions.  

The four unambiguous estimates of the intrinsic jet-disc misalignment angle (for NGC\,3557, IC\,4296, NGC\,383 and NGC\,3665) are in the range $30^\circ \la \theta_{\rm dj} \la 60^\circ$ (Table~\ref{tab:alignments}). 
In the remaining two cases, IC\,1531 and NGC\,7075, we cannot distinguish between the near and far sides of the discs and there is therefore an ambiguity in $\theta_{\rm dj}$.  Nevertheless, the possible $\theta_{\rm dj}$ angles occupy a similar range.

Despite the meagre statistics, we can set some useful constraints on the distribution of $\theta_{\rm dj}$.  In the case of an isotropic distribution of misalignments, as suggested by \citet{Verdoes05}, we would expect a probability distribution $p(\theta_{\rm dj}) d\theta_{\rm  dj} = \sin\theta_{\rm dj} d\theta_{\rm dj} (0 \leq \theta_{\rm dj} \leq 90^\circ)$. For the four sources with unambiguous $\theta_{\rm dj}$, there is no very well-aligned case, but there is a marginally significant lack of large $\theta_{\rm dj}$ (i.e.\,$\ga 60^\circ$) and the distribution is inconsistent with isotropy at the 88\% level using a Kolmogorov-Smirnov test.  A distribution in which the misalignment angles are distributed homogeneously over "polar caps" of $0^\circ \la \theta_{\rm dj} \la 60^\circ - 75^\circ$ \citep{Schmitt02} is fully consistent with our results.  Subject to confirmation from a larger sample, we conclude that jets tend to avoid the disc plane but are otherwise not preferentially aligned, in broad agreement with the statistical results of \citet{Schmitt02} and \citet{Verdoes05}. The much stronger tendency to alignment between dust discs and jet axes found in some earlier work (e.g.\ \citealt{KE79,deKoff00,deRuiter02}) must then be attributed to a combination of the effects of projection and the inclusion of a separate population of radio galaxies with irregular dust lanes, for which the alignment appears to be much closer \citep{Verdoes05}. 

Although the issue of jets misalignment with respect to the rotation axis of either the inner accretion disc or the larger-scale dust/molecular gas disc has been discussed extensively in the literature, both from the observational (e.g.\ \citealt{Schmitt01,Schmitt02,Verdoes05,Gallimore06}) and theoretical (e.g.\ \citealt{Kinney00,King07,King15,King18}) points of view, the physical mechanism causing such a wide distribution of misalignment angles is still debated. A discussion of this issue was presented in Paper~I (Section~6.5) and  only a brief summary is given below. Various scenarios have been proposed, depending (among other factors) on the physical scale of the misalignment. 

If the misalignment occurs between the jet axis and the dust/molecular gas disc on large ($\sim$kpc) scales, then misaligned external accretion of cold gas (from mergers or interactions) could be an explanation. If the externally-acquired gas is still in the process of settling into the host galaxy potential (e.g.\ \citealt{Lauer05,Shabala12,Voort15,Voort18}), its angular momentum vector is likely not to be aligned with that of the central SMBH. Low-level perturbations in the CO morphology and/or kinematics are ubiquitous in our southern radio galaxy sample (Paper~II). This, along with the presence of (possibly interacting) companions, provides supporting evidence to this scenario at least for some objects (including NGC\,3100; see Section~\ref{sec:ngc3100_VLA_discuss}). In other cases (including NGC\,3557, the source in which we measure the largest jet-disc misalignment; Table~\ref{tab:alignments}), alignment between the kinematics of gas and stars (compatible with an internal origin for the gas) and the persistence of the jet direction over the lifetime of the radio source argue against an external gas accretion (Paper~I, II).   

If the misalignment occurs between the inner edge of the observed dust/gas disc and the jet formation scale, it is possible that the jet is launched along the spin axis of the central black hole, which in turn is determined by earlier merger events and is not aligned with an axis of the stellar gravitational potential into which the gas has settled. If the jet direction is instead defined by an inner accretion disc, then warping of the disc may cause misalignment \citep{Schmitt02}. The inner part of a tilted thin accretion disc is expected to become aligned  with the black hole mid-plane via the Bardeen-Petterson effect \citep{BP75}, although the hole spin eventually becomes parallel to the angular momentum vector of the accreted matter \citep{Rees78,SF96}. Simulations  by \citet{Liska18,Liska19,Liska20} show that jets can become parallel to the angular momentum vector of the outer tilted disc in the misaligned case. Whether this mechanism works for the geometrically thick accretion discs thought to occur in LERGs is still a matter of debate \citep[e.g.][]{Zhuravlev14}.

Constraints on the relative alignment between discs on intermediate scales ($\approx 0.03  - 1$\,pc) and jets are provided by observations of H$_2$O megamaser galaxies \citep{Greene13,Kamali19}. These show that the relative orientations of maser disc rotation axes and jets in projection on the sky are usually within $\Delta \la 30^\circ$, whereas the maser discs show no tendency to align with circumnuclear structures on $<500$\,pc scales.  We note, however, that megamaser discs are typically found in spiral galaxies, often with high-excitation optical spectra, in contrast to the early-type LERGs that are the targets of our study.

\subsection{Jet-disc interactions}\label{sec:interactions}

Although there has been substantial recent work on modelling of jet-ISM interactions \citep[e.g.][]{Wagner12,Wagner16,Mukherjee16,Mukherjee18b,Cielo18}, it is not clear whether this applies to classical FR\,I jets in LERGs. The reasons are as follows.
\begin{enumerate}
\item Currently available models and simulations indicate that coupling is most pronounced if the jet propagates close to the plane of the disc, for the obvious reason that the amount of material along the jet path is maximised.  In that case, there are substantial effects on both the gas kinematics (outflows with velocities up to $\sim1000$~km~s$^{-1}$, high radial acceleration and gas turbulence) and the subsequent evolution of the jet flow (deflecting and decelerating the flow to sub-relativistic speeds).  Jets propagating closer to the disc axis are much less affected. We have established that FR\,I jets tend to avoid the  disc plane (Section~\ref{sec:thetadj}), so interactions are likely to be relatively mild.  
\item Modelling of jet-ISM interplay has so far mostly focused  on jets with powers $\ga 10^{44}$~erg~s$^{-1}$.  The interplay between the radio jets and the surrounding gas in classical low-luminosity FR\,I radio galaxies such as those analysed in this work (with typical jet powers $\la 10^{44}$~erg~s$^{-1}$; \citealp[e.g.][]{Godfrey13}) has not been explored in as much detail.
\item The physical conditions assumed for the ISM in these models are probably not representative of our sources: they assume a complex, multi-phase gaseous environment (rich in ionised as well as neutral gas) which is similar to that typically observed in very gas-rich, Seyfert-like systems with radiatively efficient accretion, as opposed to the early-type hosts of LERGs \citep[e.g.][]{Best12}.
\item Finally, the simulations mostly predict the impact of radio jets during the early phases of their evolution (i.e.\,within few million years from their first turning on; \citealp[e.g.][]{Mukherjee18b}). The majority of our sources are instead mature RGs, with evolved large-scale jets (see Paper I). 
\end{enumerate}

Despite all of these caveats, we might expect to see some signs of jet-disc interaction in the cases where the jets are closest to the disc plane. NGC\,3557 is the source for which we measure the largest secure jet-disc misalignment ($\theta_{\rm dj}=57^{\circ}\pm2^{\circ}$). In Paper II, we found clear evidence of disturbed CO kinematics in this source, possibly to be ascribed to a jet-ISM interaction (such an interaction was also tentatively claimed by \citealt{Vilaro19}, based on molecular line ratios). However, although  deviations from circular rotation are clearly visible both in the gas rotation pattern and in the velocity curve at a position that (at least in projection) appears located along the direction of the jet axis (see Paper II), these are low-level perturbations, not consistent with the high velocity outflows predicted by simulations. Neither are significant perturbations observed in the structure of the inner jet flow, which appears straight and smooth (Figure~\ref{fig:ngc3557_VLA}). If a jet-ISM interaction is occurring, then it does not appear to have much effect on the evolution of the system (at least at the spatial resolution and sensitivity of our current observations).

\citet{Morganti20b} have suggested that the impact of the jet on the surrounding medium decreases as the source grows, with the central gas becoming less kinematically perturbed as the radio jets evolve. In this regard, the numerical simulations predict that most of the (sub-)kpc scale gas affected by the transit of expanding radio jets eventually falls back into the central regions, settling into the galaxy potential on estimated timescale of the order of tens of Myr \citep[e.g.][]{Mukherjee16}. NGC~3557 may then be a case where we are observing the residual effects of an evolved interaction, although the lack of appropriate models and higher resolution molecular gas observations prevent us from drawing solid conclusions.

Features in the CO morphology/kinematics like those described for NGC\,3557 are also observed in IC\,1531, possibly suggesting again a jet-ISM interaction (see Paper II). However, the many uncertainties affecting the analysis of both the molecular gas and the radio jet structure of this source (see Table~\ref{tab:alignments} and Paper I) do not allow us to put robust constraints on this scenario.

\subsection{The case of NGC 3100}\label{sec:ngc3100_VLA_discuss}
\begin{figure}
\centering
\includegraphics[scale=0.33]{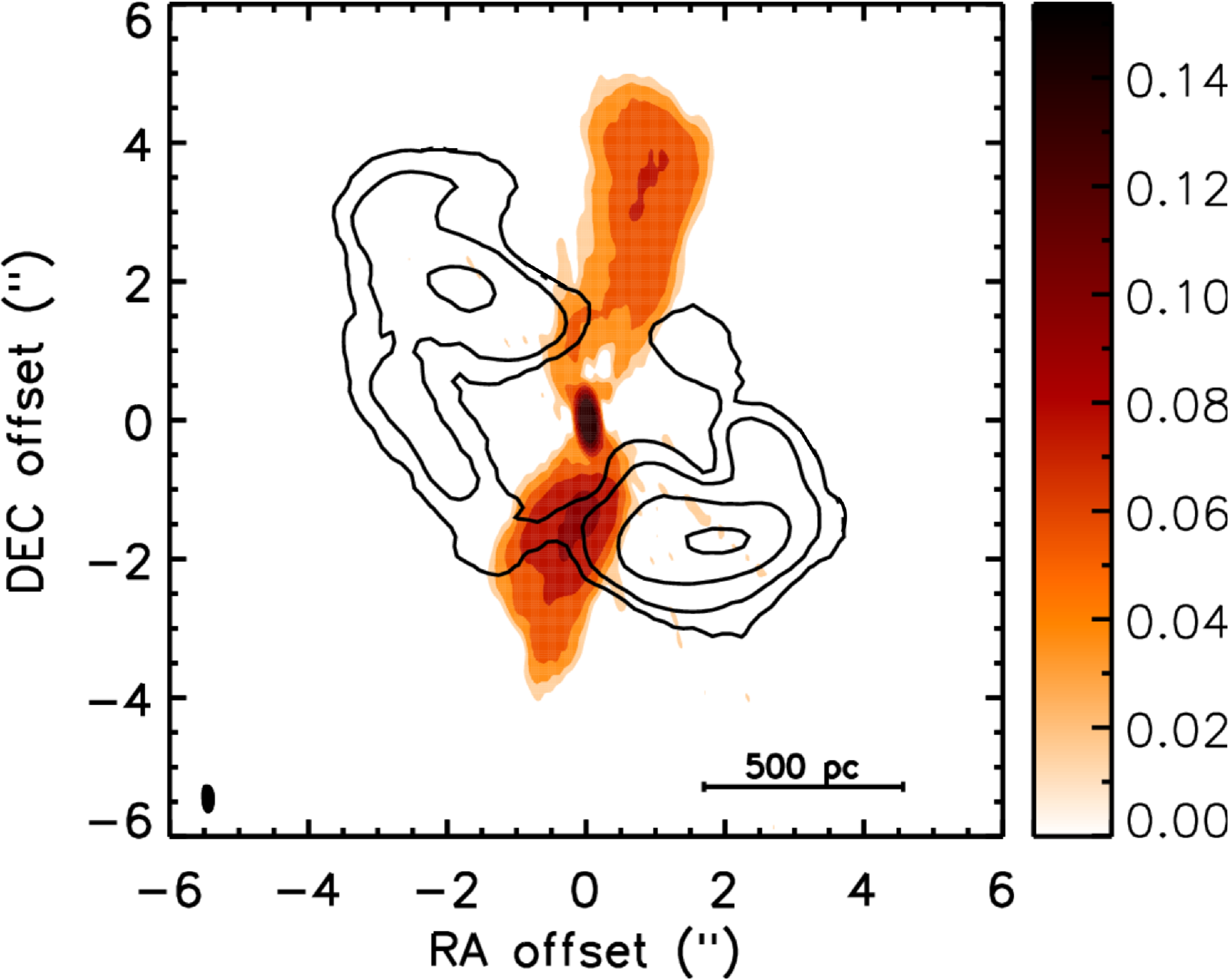}
\caption[]{\small{Masked JVLA continuum map of NGC\,3100 at 10~GHz with ALMA CO(2-1) integrated intensity contours superimposed in black. CO contours are drawn at 9, 27, 81 and 180 times the 1$\sigma$ rms noise level ($1\sigma=0.6$~mJy~beam$^{-1}$; see Paper I). The beam and the physical scale bars are drawn in the bottom-left and bottom-right corners, respectively. The wedge on the right shows the colour scale in Jy~beam$^{-1}$.}}\label{fig:NGC3100_jetDisc}
\end{figure}

The presence of a jet-ISM interaction in NGC\,3100 has been inferred from several observational signatures (see Papers I and II). The well-resolved radio structure observed in our newly-acquired 10~GHz JVLA data further strengthens this scenario. A masked (i.e.\,noise suppressed) version of the map in Panel~\ref{fig:ngc3100_VLA} is shown in Figure~\ref{fig:NGC3100_jetDisc}, with contours of the CO(2-1) emission observed with  ALMA overlaid. The properties of the jet flow in NGC\,3100 are clearly different from those observed in the other four objects  (see Figure~\ref{fig:VLAcont}) and in typical FR\,I LERGs (see Section~\ref{sec:method}). The two jets are rather asymmetric, both in surface brightness and morphology: the northern jet is fainter and also undergoes an evident bend. It is clear from Figure~\ref{fig:NGC3100_jetDisc} that the distortion of the northern jet is morphologically coincident (at least in projection) with the disruption of the CO disc (at Declination offsets between 1 and 2\,arcsec). The jet flow decollimates abruptly and decreases in surface brightness at distances of $\approx 0.8''$ ($\approx160$~pc) on both sides of the core, qualitatively consistent with rapid deceleration. There is no evidence for emission on scales significantly larger than that visible in Figure~\ref{fig:NGC3100_jetDisc} either in archival VLA data (Paper~I) or in low-resolution images at 843~MHz from the Sydney University Molonglo Sky Survey (SUMSS; \citealp{Mauch03}). The total extent of the radio source is therefore $\approx 2$\,kpc.  Based on the well-known size-age correlation of radio galaxies \citep[e.g.][]{Fanti95,ODea98}, NGC~3100 is likely to be in an early phase of its evolution. It also has the lowest monochromatic luminosity of any of the sources in our complete sample ($P_{\rm 1.4GHz} = 10^{23.0}$\,W\,Hz$^{-1}$). All of these considerations support a scenario in which we are observing a young, weak radio galaxy, whose jets are close to the disc plane.  

Nevertheless, we only observe low-amplitude kinematic perturbations in NGC\,3100, in the form of non-circular motions with peak velocities around $\pm40$~km~s$^{-1}$ (see Paper II). Such features do not seem to be consistent with the strong outflows predicted by jet feedback models \citep[e.g.][]{Wagner12,Wagner16,Mukherjee18b}. Our results for NGC\,3100 therefore suggest a much less extreme interaction between the radio jets and the surrounding ISM.  There is little evidence from the form of the CO velocity field for a jet-driven molecular outflow, strong shocks (which would significantly increase the velocity dispersion) or compression of cool gas in the disc leading to enhanced star formation. Instead,
the jets disrupt and decollimate on a scale of $\approx 0.8''$ ($\approx 160$\,pc) and create gaps in the disc at larger distances. Their effect on the molecular gas is local, rather than global (Paper~II).  NGC\,3100 appears, in some respects, to be an intermediate case between the classical FR\,I jet sources and extreme jet-gas interactions like IC\,5063 \citep{Mukherjee18a}. In the former class, entrainment is thought to occur at a very low rate, predominantly from the hot phase of the ISM and mass loss from stars embedded in the jet. The jets are then able to recollimate in the steep external pressure gradient on kpc scales and propagate to large distances \citep[e.g.][]{Laing02b,Perucho19}. In
NGC\,3100, the jets are weak and may not yet have reached a region with a steep pressure gradient before undergoing significant entrainment.  Simulations of weak, low-density jets propagating in a galaxy core of roughly constant density \citep[e.g.][]{Rossi2020} may
therefore be most relevant to NGC\,3100.

A detailed analysis of the physics of the interaction process in this object, based on multiple molecular line transitions (observed during ALMA Cycle 6) and ad hoc modelling of the jet-ISM interplay will be presented in forthcoming papers.

\section{Summary and conclusions}\label{sec:conclusion}
This is the third paper of a series aiming to investigate the multi-phase properties of a complete, volume- and flux-limited ($z<0.03$, S\textsubscript{2.7 GHz}$\geq0.25$~Jy) sample of eleven LERGs in the southern sky.

Here we have presented new, deep, high-resolution JVLA observations at 10~GHz for a subset of five sources. One source, NGC\,3100, has a distorted radio structure with clear evidence for jet-ISM interactions. The remaining four have classical FR\,I jet structures and our new images have been used to derive estimates of the jet inclination to the line of sight. Combining the newly-acquired data with previous ALMA CO observations and optical (HST) imaging from which the associated dust component is inferred, we have performed a full 3D analysis of the relative orientations of the radio jets and gas discs ($\theta_{\rm dj}$). We also added in our analysis two other FR\,I LERGs with comparable data from the literature. Full information is available for four out of six sources; the other two lack certain indications on the near/far side of the CO disc and thus the solution for $\theta_{\rm dj}$ is ambiguous. Although the number of sources is small, this is the first time this method has been used and it can be extended to much larger samples in the future.

Our main results can be summarised as follows.
\begin{itemize}
\item Misalignment angles are typically in the range $30^\circ \la \theta_{\rm dj} \la 60^\circ$.  There is no secure case showing precise alignment ($\theta_{\rm dj} < 30^\circ$). The distribution of $\theta_{\rm dj}$ is also marginally inconsistent with isotropy, in the sense that there is a tendency for jets to avoid the disc plane, as found previously by \citet{Schmitt02}.
\item The largest jet-disc misalignment ($\theta_{\rm dj}=57^{\circ}\pm2^{\circ}$) is observed in  NGC\,3557, which shows kinematic evidence for a jet-disc interaction. This is qualitatively in agreement with expectations (stronger interactions are expected for jets close to the disc plane), but far less extreme than predicted by current simulations.  It is likely that these simulations model jets which are much more powerful and younger than those in our sample and which propagate in more extreme gaseous environments.
\item In NGC\,3100, the low radio luminosity and compact size of the radio (core $+$ jets) source ($\approx2$~kpc in total), the asymmetries in both the morphology and surface brightness of the two jets, and the observed disruption of the CO disc at a bend in the northern jet, are consistent with young, weak radio jets propagating close to the disc plane and rapidly decelerating. This provides support to the scenario of a jet-ISM interaction already inferred for this source. However, the lack of extreme kinematic perturbations in the CO velocity field seems to suggest that the transit of the expanding radio jets produces only local, rather than global, effects, making NGC\,3100 as an intermediate case between classical FR\,I jet sources and extreme cases of jet-ISM interactions.
\end{itemize} 

Our analysis adds to the developing picture of LERGs as unexpectedly complex systems. We have already  demonstrated that they contain surprisingly large masses of molecular gas, despite the low accretion rates onto their central black holes (Paper~I). The gas appears to be in roughly ordered rotation, but with subtle deviations which could be attributed either to interactions with the jets  or to settling effects (Paper~II). Our latest results confirm that there is no simple relation between the rotation axis of the gas and the axis of the radio jets (by implication defined by the axis of the inner accretion disc and/or the spin of the black hole).   Further work using multiple molecular lines and tracers of  stellar and ionised gas dynamics will be required to unravel the details of jet-gas interactions, fuelling and feedback in these systems.

\section*{Acknowledgements}
The authors thank the referee, Dr.\,Santiago Garcia-Burillo, for constructive comments that helped us improving the original manuscript.
This work was partially supported by the italian space agency (ASI) through the grant "Attivit\'{a} di Studio per la comunit\'{a} scientifica di astrofisica delle alte energie e fisica astroparticellare" (Accordo Attuativo ASI-INAF n.2017-14-H.0). IP and IR acknowledge support from PRIN MIUR project "Black Hole winds and the Baryon Life Cycle of Galaxies: the stone-guest at the galaxy evolution supper”, contract \#2017PH3WAT. IP acknowledges support from INAF under the PRIN SKA project FORECaST.
This paper makes use of the following ALMA data: ADS/JAO.ALMA\#[2015.1.01572.S]. ALMA is a partnership of ESO (representing its member states), NSF (USA) and NINS (Japan), together with NRC (Canada), NSC and ASIAA (Taiwan), and KASI (Republic of Korea), in cooperation with the Republic of Chile. The Joint ALMA Observatory is operated by ESO, AUI/NRAO and NAOJ. The National Radio Astronomy Observatory is a facility of the National Science Foundation operated under cooperative agreement by Associated Universities, Inc. This paper has also made use of the NASA/IPAC Extragalactic Database (NED) which is operated by the Jet Propulsion Laboratory, California Institute of Technology under contract with NASA.

\section*{Data availability}
The raw JVLA data used in this article are available to download at the VLA archive website (https://science.nrao.edu/facilities/vla/archive/index; project code: 18A-200, PI: I.\,Ruffa). The ALMA data can be downloaded from the ALMA archive query (https://almascience.nrao.edu/asax/; project code: 2015.1.01572.S, PI: I.\,Prandoni). The calibrated data, final data products and original plots generated for the research study underlying this article will be shared upon reasonable request to the first author.




\bibliographystyle{mnras}
\bibliography{mybibliography} 




\appendix
\section{Geometry of the jet-CO disc system in 3D}\label{sec:Appendix}

\begin{figure}
\centering
\includegraphics[scale=0.65]{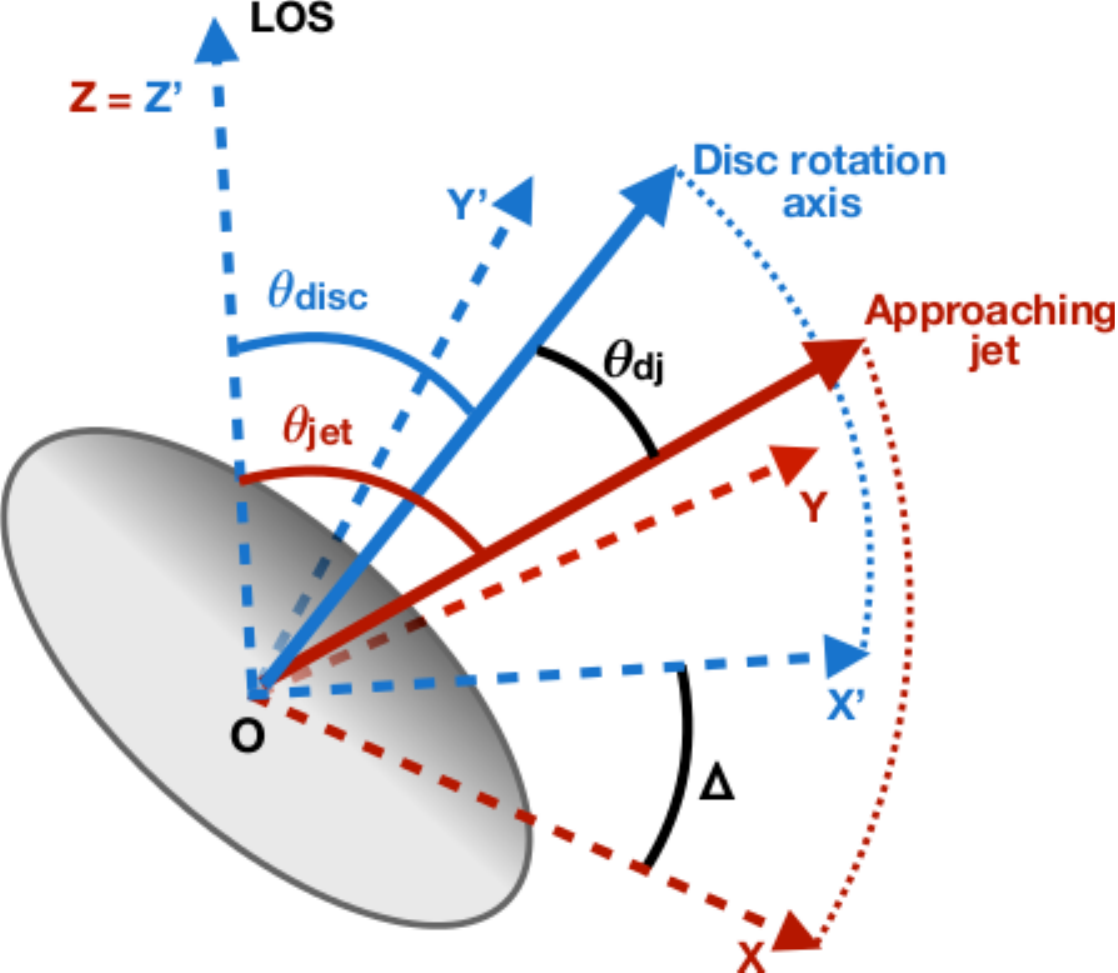}
\caption[3D geometry of the jet and the CO disc]{\small{Sketch describing the 3D geometry of the jet-disc system for an illustrative case. The Cartesian system O$xyz$ is built so that the origin is in the core, the $z$ axis is along the line of sight (positive sign towards the observer), and the $x$-$y$ axes are in the plane of the sky. The $x$ axis is parallel to the projection of the jet (positive sign along the projection of the approaching jet, indicated as a solid red arrow). The angle between the approaching jet and the line of sight is $\theta_{\rm jet}$. Similarly, in the O$x^\prime y^\prime z^\prime$ Cartesian system, the origin is in the core, the $z^\prime$ axis is again along the line of sight, the $x^\prime$-$y^\prime$ axes are in the plane of the sky, and $x^\prime$ lies along the projection of the disc rotation axis (i.e.\,the vector normal to the disc plane, indicated as a solid blue arrow). The inclination of the gas disc to the line of sight is given by the angle $\theta_{\rm disc}$. $\theta_{\rm dj}$ is the angle between the jet and the disc rotation axes and defines the intrinsic relative orientation of the CO disc and the radio jets. The {\em projected} jet-disc relative orientation is given by the angle $\Delta$. The grey ellipse illustrates the CO disc, a darker grey shade is used to indicate the near side of the disc. See the text for details.}}\label{fig:jet_disc_geometry}
\end{figure}

Figure~\ref{fig:jet_disc_geometry} illustrates the model parameters and coordinate systems used to describe the jet-disc geometry in 3D.
We use a right-handed Cartesian coordinate system $xyz$, such that the origin is in the core, the $z$ axis is along the line-of-sight (positive sign towards the observer) and the $x$ and $y$ axes are in the plane of the sky. The $x$ axis is parallel to the projection of the jet axis (positive sign along the projection of the approaching jet). The angle between the approaching jet and the line of sight (i.e.\,the jet inclination) is $\theta_{\rm jet}$ (ranged $0 \leq \theta_{\rm jet} \leq \pi/2$).
In this system, a unit vector along the
approaching jet is

\begin{eqnarray*}
\begin{pmatrix}
+\sin \theta_{\rm jet} \\ 0 \\ +\cos \theta_{\rm jet}
\end{pmatrix}\, .
\end{eqnarray*}

Similarly, we define a second Cartesian coordinate system, $x^\prime y^\prime z^\prime$, in which $z^\prime$ is again along the line-of-sight, $x^\prime$ and $y^\prime$ are in the plane of the sky, but $x^\prime$ is along the projection of the disc rotation axis (i.e.\,the vector normal to the disc plane).  $\theta_{\rm disc}$ is the angle between the disc rotation axis and the line of sight (i.e.\,the inclination of the disc, ranged $0 \leq \theta_{\rm disc} \leq \pi/2$). The unit vector components of the disc rotation axis in the primed coordinate system are

\begin{eqnarray*}
\begin{pmatrix}
+\sin \theta_{\rm disc} \\ 0 \\ +\cos \theta_{\rm disc} 
\end{pmatrix} \, .
\end{eqnarray*}

The two coordinate systems are related by a rotation of an angle $\Delta$ about the common $z$ ($z^\prime$) axis. In our notation, this is the relative orientation of the jet and disc rotation axis in projection, i.e.\,$\Delta = |PA_{\rm rot} - PA_{\rm jet}|$, where $PA_{\rm rot}$ and $PA_{\rm jet}$ are the position angles of the CO disc rotation axis and the approaching radio jet, respectively (see Section~\ref{sec:thetadj} and Table~\ref{tab:alignments} for details). The sign of the rotation does not affect the final result of this calculation, so we range $\Delta$ in $[0,\pi]$: if $\Delta < \pi/2$, then the near side of the disc appears in projection on the receding jet side; if $\Delta > \pi/2$, then it is projected on the approaching jet.
 
The components of a vector in the $xyz$ frame are related to the components in the $x^\prime y^\prime z^\prime$ frame by

\begin{eqnarray*}
  \begin{pmatrix}
    x \\ y \\ z
  \end{pmatrix}
  =
  \begin{pmatrix}
    +\cos\Delta & +\sin\Delta & 0 \\
    -\sin\Delta & +\cos\Delta & 0 \\
    0 & 0 & 1\\
  \end{pmatrix}
  \begin{pmatrix}
    x^\prime \\ y^\prime \\ z^\prime
  \end{pmatrix}\, .
\end{eqnarray*}

The components of the disc rotation axis vector in the $xyz$ frame are therefore

\begin{eqnarray*}
  \begin{pmatrix}
    +\cos\Delta & +\sin\Delta & 0 \\
    -\sin\Delta & +\cos\Delta & 0 \\
    0 & 0 & 1\\
  \end{pmatrix}
  \begin{pmatrix}
+\sin \theta_{\rm disc} \\ 0 \\ +\cos \theta_{\rm disc} 
\end{pmatrix}
  =
  \begin{pmatrix}
    +\cos\Delta \sin \theta_{\rm disc} \\
    -\sin\Delta \sin \theta_{\rm disc} \\
    +\cos \theta_{\rm disc} \\
  \end{pmatrix} \, .
\end{eqnarray*}

The angle between the jet and disc rotation axis vectors, $\theta_{\rm dj}$, is obtained by taking the dot product of the two unit vectors:
\begin{eqnarray*}
  \begin{pmatrix}
    +\cos\Delta \sin \theta_{\rm disc} \\
    -\sin\sin \theta_{\rm disc} \\
    +\cos \theta_{\rm disc} \\
  \end{pmatrix}
  .
\begin{pmatrix}
+\sin \theta_{\rm jet} \\ 0 \\ +\cos \theta_{\rm jet} 
\end{pmatrix}
= \cos\Delta \sin \theta_{\rm disc}\sin \theta_{\rm jet} + \cos \theta_{\rm disc}\cos \theta_{\rm jet} 
\end{eqnarray*}
so
\begin{eqnarray}\label{eq:inclination_angleA1}  
  \cos\theta_{\rm dj} = |\cos\Delta \sin \theta_{\rm disc}\sin \theta_{\rm jet} + \cos \theta_{\rm disc}\cos \theta_{\rm jet}|
\end{eqnarray}
(modulus because we want to range in $[0, \pi/2]$). 



\bsp	
\label{lastpage}
\end{document}